\documentclass[prd,twocolumn,superscriptaddress,showpacs,longbibliography,showkeys]{revtex4-2}
\usepackage[english]{babel} 
\usepackage{booktabs} 
\usepackage{enumitem} 
\setlist[itemize]{noitemsep} 
\usepackage[normalem]{ulem}

\usepackage[dvipsnames,usenames]{xcolor}
\usepackage{braket}
\usepackage{amsfonts}
\usepackage{amsmath}
\usepackage{amssymb}
\usepackage{mathtools}
\usepackage{bbold}
\usepackage{float}
\usepackage{physics}
\usepackage{tikz}
\usepackage{hyperref}
\hypersetup{colorlinks,linkcolor={blue},citecolor={green},urlcolor={blue}}
\usepackage{qcircuit}
\usepackage{ulem}

\begin{document}

\title{Trapped-Ion Quantum Simulation of Collective Neutrino Oscillations}

\newcommand{\UNITN}{{Dipartimento di Fisica, University of Trento, via Sommarive 14, I–38123, Povo, Trento, Italy}}
\newcommand{\TIFPA}{INFN-TIFPA Trento Institute of Fundamental Physics and Applications,  Trento, Italy}
\newcommand{\ECT}{ECT*, European Center for Theoretical Studies in Nuclear Physics and Related Areas, Strada delle Tabarelle 286, Trento, Italy}

\author{Valentina Amitrano}
\affiliation{\UNITN}
\affiliation{\TIFPA}
\author{Alessandro Roggero}
\affiliation{\UNITN}
\affiliation{\TIFPA}
\author{Piero Luchi}
\affiliation{\UNITN}
\affiliation{\TIFPA}
\author{Francesco Turro}
\affiliation{\UNITN}
\affiliation{\TIFPA}
\author{Luca Vespucci}
\affiliation{\UNITN}
\affiliation{\TIFPA}
\affiliation{\ECT}
\author{Francesco Pederiva}
\affiliation{\UNITN}
\affiliation{\TIFPA}

\date{\today}

\begin{abstract}
It is well known that the neutrino flavor in extreme astrophysical environments changes under the effect of three contributions: the vacuum oscillation, the interaction with the surrounding matter, and the collective oscillations due to interactions between different neutrinos. The latter adds a nonlinear contribution to the equations of motion, making the description of their dynamics complex.
In this work we study various strategies to simulate the coherent collective oscillations of a system of $N$ neutrinos in the two-flavor approximation using quantum computation. This was achieved by using a pair-neutrino decomposition designed to account for the fact that the flavor Hamiltonian, in the presence of the neutrino-neutrino term, presents an all-to-all interaction that makes the implementation of the evolution dependent on the qubit topology.

We analyze the Trotter error caused by the decomposition demonstrating that the complexity of the implementation of time evolution scales polynomially with the number of neutrinos
and that the noise from near-term quantum device simulation can be reduced by optimizing the quantum circuit decomposition and exploiting a full-qubit connectivity. We find that the gate complexity using second order Trotter-Suzuki formulas scales better with system size than with other decomposition methods such as Quantum Signal Processing.
We finally present the application and the results of our algorithm on a real quantum device based on trapped-ion qubits.
\end{abstract}

\maketitle

\section{Introduction}
\label{Section_introduction}
Quantum computation can provide an enormous advantage for the physical description of many-body quantum systems due to the fact that it does not necessarily require an exponential scaling of computational resources as the size of the system increases~\cite{Feynman1982,Lloyd1073}. 
The prospect of employing quantum devices to study standard model physics has led to a world wide effort to design algorithms and apply them to currently available quantum platforms (see, e.g., Ref.~\cite{klco2022standard} for a recent review).

Given a quantum system in some fixed initial state $\ket {\Psi_0} $, its time evolution under the Hamiltonian $ H $ is given by the action of the real-time evolution operator:
\begin{equation}
   U(t) = e^{-iHt} \,,
   \label{U}
\end{equation}
transforming the state according to the time-dependent Schr\"odinger equation $\ket{\Psi(t)} = U(t) \ket{\Psi_0}$. In general, a direct approach based on this description faces an exponentially growing cost on classical computers as the size of the system increases due to both the 
enormous memory requirements to encode the states of the system
and to the operational cost needed to perform matrix multiplications. Important exceptions to this behavior are found, for instance, in stabilizer states~\cite{Gottesman:1997zz,Gottesman:1998hu} or in systems with low levels of bipartite entanglement~\cite{PhysRevLett.91.147902,PhysRevLett.100.030504}.

According to the Deutsch model~\cite{deutsch1985} of a quantum computer, given a system in a pure state $ \ket{\varphi_0}$, a quantum algorithm consists of a unitary transformation $U$ which produces a certain final state $\ket{\varphi_f} = U \ket{\varphi_0} $ according to quantum mechanical rules. It is then possible to perform quantum measurements yielding the probability of finding the systems in a given state of a given basis, probability that constitutes the result of the calculation. 
The Solovay–Kitaev theorem \cite{nielsen2001quantum} demonstrates that there is a finite set of quantum gates, which can approximate, with arbitrary accuracy, any unitary transformation $U$. In this sense, the Deutsch model is universal.

\begin{figure}
    \centering
    \includegraphics[width = 8cm]{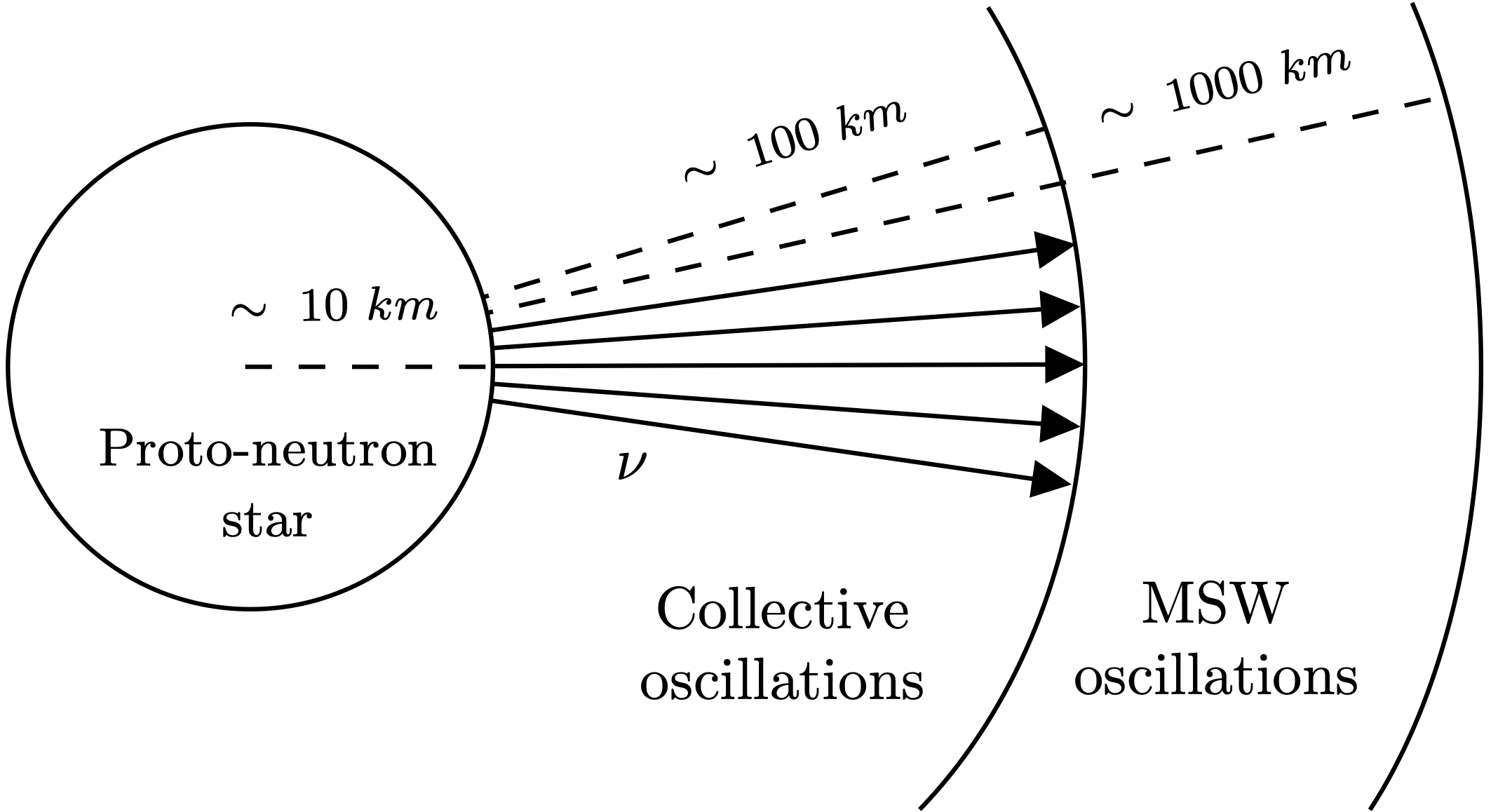}
    \caption{Sketch of the environment in a core-collapse supernova assuming spherical symmetry. Near the proto-neutron star the large local neutrino density causes an important effect from pairwise neutrinos scattering, while at large radii the interaction with electrons prevails instead. Moreover, we assume a narrow cone of forward peaked emitted neutrinos.}
    \label{Supernovae}
\end{figure}

An interesting many-body system amenable for exploring simulations of the time evolution on a quantum computer is that of collective flavor oscillations of neutrinos caused by forward neutrino-neutrino scattering. These are predicted to occur in extreme astrophysical environments like core-collapse supernovae, neutron star mergers, and the early Universe \cite{PANTALEONE1992,Pantaleone92,qian1995neutrino,PhysRevD.53.5382,Pastor2002B,Balantekin_2005,PhysRevD.95.103007}.
The description of flavor oscillations is a crucial aspect of such studies since the physics of matter under extreme conditions is strongly flavor dependent \cite{qian1995matter, qian1993connection}, and moreover the energy spectrum is different for different neutrino flavors \cite{Janka2012}.
A star with a sufficiently large mass ($ \gtrapprox 8\, \text{M}_{\odot} $) undergoes a gravitational collapse that can result in a neutron star or a black hole. During the collapse it emits a very large amount of energy ($ \sim 10^{53} $ erg) in the form of a large number of neutrinos ($ \sim 10^{58} $). In a few seconds the $ 10 \% $ of the gravitational mass of the star is converted into neutrinos flowing with an energy $E_\nu \simeq (10 \divisionsymbol 30)$ MeV.
The evolution of this neutrino sea plays a fundamental role in supernova collapse phenomena. They are in fact the main carriers of the lepton number within the reactions taking place in the interior. Moreover, they are responsible for the loss of entropy and can increase the instability of the star by eventually generating the explosion \cite{duan2010review}.
A simple diagram showing the regions where different neutrino processes are active in a core collapse supernovae, assumed to have spherical symmetry, is shown in Fig.~\ref{Supernovae}. Collective neutrino oscillations are generally expected to be dominant in a range of intermediate distances from the core ($ \sim 100 $ km) where the density of neutrinos is large while the external lepton electron density is not sufficiently large to suppress flavor oscillations~\cite{duan2010review}. The external shell, in which the neutrino density is lower, is instead dominated by vacuum oscillations and interactions with the surrounding matter leading to the MSW effect~\cite{smirnov2005msw}. 

A full description of the dynamical evolution of flavor in these processes is hindered by the large computational cost required to carry out simulations with large numbers of interacting neutrinos. A common approach adopted to circumvent
the problem is to use a mean-field approximation for the equation of motions, allowing one to study large-scale systems with complex geometries~\cite{pantaleone1992neutrino, duan2006coherent, duan2006simulations, chakraborty2016collective}.
A full treatment of correlation effects in the complete many-body evolution can, however, be attained in relatively small systems with $O(10)$ neutrinos~\cite{Cervia2019,Rrapaj2020,Patwardhan2021}, in situations of large symmetry~\cite{Birol2018,Martin2022,Xiong2022,roggero2022entanglement} and/or small levels of bipartite entanglement using tensor network methods~\cite{RoggeroMPS2021,RoggeroDPT2021,cervia2022collective}. 
Semiclassical methods are another class of approaches that retain some of the correlations while maintaining numerical efficiency; these methods were recently applied to the neutrino problem in Ref.~\cite{lacroix2022}.
Quantum simulations offer an alternative to explore out-of-equilibrium flavor dynamics in regimes that are not accessible by these classical approaches. 
Early calculations on small systems with up to four neutrinos have been carried out on both digital quantum computers~\cite{hall2021simulation,yeter2022collective} and quantum annealers~\cite{illa2022basic} showing that both careful algorithm design and error mitigation techniques have to be considered when tackling this challenging problem on current generation devices.
In this work we propose an efficient quantum algorithm to describe the evolution of the flavor state of a many-neutrino system using a digital quantum computer and paying attention to the optimal quantum gate decomposition and to the complexity of the quantum circuit needed for the simulation. We analyze in detail the scaling of the Trotter error, the number of operations needed to perform the evolution at a fixed error, and the complexity of the quantum gate decomposition of the evolution operator, paying attention to the machine-aware compilation that has to consider the topology of the qubit system.
We note that to obtain physical information about the many-neutrino system one does not need to simulate the evolution of all the $10^{58}$ emitted neutrinos, but it is sufficient to limit the simulation only to a number of neutrinos in a space region that is causally connected. Furthermore, collective oscillations can operate on timescales much smaller than the total propagation time needed for a neutrino to leave the system starting from the neutrino sphere. Finally, even small-scale simulations can provide crucial information about the conditions required for specific collective modes to appear as well as the nature of correlations responsible for them (see, e.g., Refs.~\cite{RoggeroMPS2021,RoggeroDPT2021,Patwardhan2021,roggero2022entanglement,Martin2022}).

In Sec.~\ref{Sec2} we present the description of the physical system of several neutrinos used in this work. In Sec.~\ref{Sec3} we analyze the decomposition of the unitary propagator in Eq.~\eqref{U} containing an all-to-all interaction. In order to do that, we exploit the pair property of the potential, showing some of the advantages of a full-qubit connectivity which guarantees greater freedom in the decomposition and therefore less complexity of the final implementation.
In Sec.~\ref{Sec5} we present the optimal decomposition technique to find the best quantum circuit for the trapped-ion quantum device used to perform the real quantum simulation [a Quantinuum System Model (QSM) H1-2
trapped-ion device].
In Sec.~\ref{Sec6} we report the results obtained from the real quantum simulations for a single Trotter step propagation using different time steps and for multistep long time evolution in the cases of $N=4$ and $N=8$ neutrinos.
Finally, in Sec.~\ref{Sec4} we present a detailed derivation of the Trotter error scaling introduced by the decomposition of $U(t)$ and the time discretization. We also compare the gate complexity to that of other decomposition methods such as qubitization.
Finally, we demonstrate the usefulness of the decomposition proposed in Sec.~\ref{Sec3}, which ensures that the number of necessary operations follows a low degree polynomial with the number of particles. 

\section{SU(2) model of a many-neutrino system}
\label{Sec2}
The Hamiltonian of a many-neutrino system including the effect of forward scattering is equivalent to an all-to-all coupled spin system and thus represents an interesting
many-body quantum problem governed by the weak interaction. 
The first approximation we make is to consider only two neutrino flavors in the description: the electron flavor $\nu_e$ and a single heavy flavor $\nu_x$ which is a combination of $\mu$ and $\tau$ neutrinos. This can be justified if the mixing angle $\theta_{13}=0$ as shown in~\cite{balantekin1999constraints}. 
In this way the general flavor state of each neutrino, given by the two-flavor superposition $\ket{\Psi} = \alpha \ket{\nu_e} + \beta \ket{\nu_x}$, can be fully encoded in the state of a qubit by means of the mapping:
\begin{equation}
    \ket{\nu_{e}} \longmapsto \ket{0}, \quad \ket{\nu_{x}} \longmapsto \ket{1}\,.
    \label{encoding}
\end{equation}
The flavor Hamiltonian in this basis can be decomposed into three main terms~\cite{duan2010review,pehlivan2011invariants}: (1) a one-body contribution describing vacuum mixing $ H_{vac}\coloneqq H^{(1)} $ which takes into account the flavor oscillations of each neutrino due to the misalignment between flavor states and mass eigenstates; 
(2) a second one-body part describing the coupling to external matter leading to the MSW effect 
and (3) the neutrino-neutrino interaction term $ H_{\nu \nu}: = H ^ {(2)} $ generated by forward scattering.
In this work we neglect the second contribution since we are interested in describing the dynamics in the coherent oscillation-dominated region, 
and only focus on a simplified Hamiltonian of the form~\cite{pehlivan2011invariants}
\begin{equation}
\begin{split}
    H = H^{(1)} + H^{(2)}&= \sum_{i=0}^{N-1} h_i + \sum_{i<j}^{N-1} h_{ij} \\
    &= \sum_{i = 0}^{N-1} \boldsymbol{b} \cdot \boldsymbol{\sigma}_i + \sum_{i < j}^{N-1} J_{ij} \boldsymbol{\sigma}_i \cdot \boldsymbol{\sigma}_j\;,
     \label{H}
\end{split}
\end{equation}
where we used bold symbols to denote three-dimensional vectors. The vectors $\boldsymbol{\sigma}_i=(\sigma_i^{(x)},\sigma_i^{(y)},\sigma_i^{(z)})$ are formed by the Pauli matrices acting on the $i$th neutrino. Here and in the following we suppress the identity operators acting on the other spins, for example, $\sigma^{(x)}_1 = \mathbb{1} \otimes \sigma^{(x)} \otimes \mathbb{1} \otimes  \dots \otimes \mathbb{1}$.
The simple structure of the neutrino-neutrino interaction term originates from treating neutrinos as plane waves with definite momentum and the weak interaction as a contact term in coordinate space. In order to extend this formulation to take into account, in a more realistic way, the spatial localization of neutrinos, especially important in inhomogeneous systems~\cite{Stirner_2018}, it is possible to consider instead the evolution of neutrino wave packets. Extensions along these lines are left for future work.
In the flavor basis, the vector $\boldsymbol{b}$ in the first term describes vacuum mixing of neutrinos with the same energy and is given explicitly by
\begin{equation}
    \boldsymbol{b} = \frac{\delta m^2}{4 E}( \sin(2 \theta_{\nu}), 0, - \cos(2\theta_{\nu}))\;.
\end{equation}
In this expression $\delta m^2 = m_2^2 - m_1^2 $ is the square mass difference between mass eigenstates, which is of order $10^{-4} \text{ eV}^2$, $\theta_\nu$ is the mixing angle that we took equal to $ \theta _ {\nu} = 0.195 $, and $ E $ is the neutrino energy.
The two-body interaction term is described by the $SU(2)$ invariant product of Pauli matrices in which, also in this case, the tensor products with the other particles are implicit. For example,
\begin{equation}
    \sigma^{(x)}_0 \sigma^{(x)}_2 = \sigma^{(x)} \otimes \mathbb{1} \otimes \sigma^{(x)} \otimes \mathbb{1} \otimes \dots \otimes \mathbb{1}\;.
\end{equation}
The coupling constant of the pair interaction can be written explicitly as follows,
\begin{equation}
\begin{split}
\label{eq:twobcoupling}
J_{ij} &= \frac{\sqrt{2}G_F}{V}(1 - \cos(\theta_{ij})) \coloneqq \frac{\mu}{N}(1 - \cos(\theta_{ij}))\,,
\end{split}
\end{equation}
and it depends on the relative angle of propagation,
\begin{equation}
\cos(\theta_{ij})=\frac{\boldsymbol{p}_i\cdot\boldsymbol{p}_j}{\|\boldsymbol{p}_i\|\|\boldsymbol{p}_j\|}\;,
\end{equation}
where $\boldsymbol{p}_i$ is the momentum of the $i$th neutrino.
This means that the neutrinos that interact the most are those that propagate in directions with a larger relative angle.
In Eq.~\eqref{eq:twobcoupling} we introduced the energy scale $\mu=\sqrt{2}G_F n_{\nu}$, where $G_F$ if the Fermi constant, $n_{\nu}=N/V$ the neutrino number density, and $V$ the volume of the system. 
We choose the neutrino energy in order to obtain the same coupling constant for the one-body and two-body energies,
\begin{equation}
    \frac{\mu}{N} = \frac{\delta m^2}{4 E}\;,
\end{equation}
and measure time in units of $\mu^{-1}$.
\begin{figure}
    \centering
    \includegraphics[width = 8.5cm]{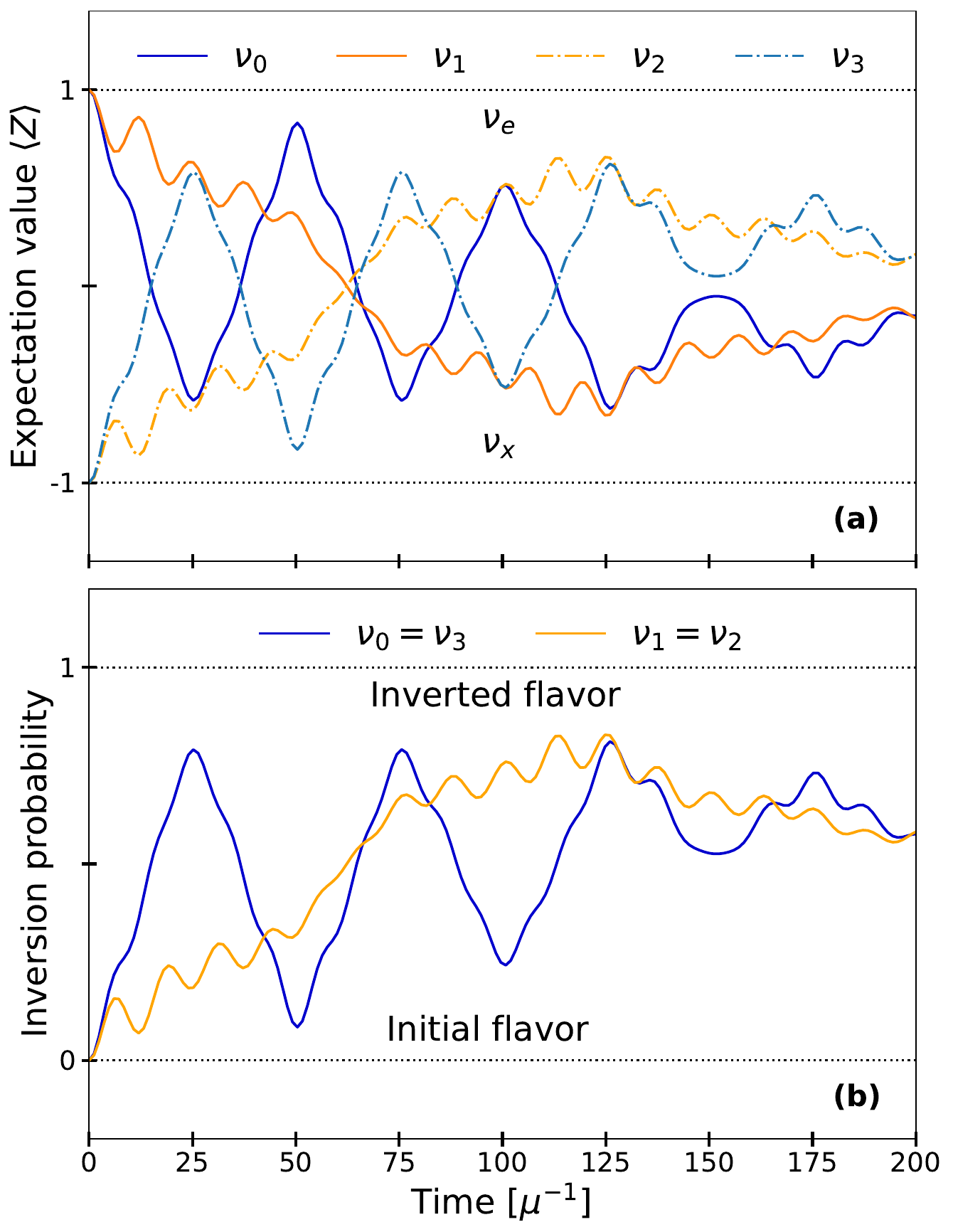}
    \caption{Exact evolution of a system of $ N = 4 $ neutrinos with initial state $ \ket{\Psi_0}=\ket {0011} $. The panel (a) shows the evolution of the expectation value $ \expval{Z_i} $ while the panel (b) shows the evolution of the flavor inversion probability.}
    \label{Th_evo}
\end{figure}
As done in previous work~\cite{hall2021simulation}, we take a simple grid of angles,
\begin{equation}
    \theta_{ij} = \arccos(0.9) \frac{\abs{i-j}}{(N-1)}\,,
    \label{theta_def}
\end{equation}
meant to reproduce a narrow cone of forward peaked neutrinos in accordance with the geometry displayed in Fig.~\ref{Supernovae}.
With this choice of angular distribution, and for even $N$, the neutrino Hamiltonian in Eq.~\eqref{H} turns out to be symmetric under the particle exchange
\begin{equation}
\begin{split}
    \nu_k &\longleftrightarrow \nu_{N-1-k}\;,
        \label{Eq_symm_exchange}
\end{split}
\end{equation}
for $k=0,\dots,N/2$.
The initial state of the system, used for all the simulations presented in this work, consists of setting the first $ N/2 $ neutrinos in the $ \ket {\nu_e} $ flavor state and the other $ N/2 $ in the $ \ket{\nu_{x}}$ state. In this way the initial state is symmetric under the composition of particle exchange and flavor inversion. For $N = 4$ neutrinos, for instance, the initial state is given by
\begin{equation}
\label{eq:psi0_4}
    \ket{\Psi_0} = \ket{\nu_0 \nu_1 \nu_2 \nu_3} = \ket{0011}=\ket{\nu_e \nu_e \nu_x \nu_x}\;.
\end{equation}
One can obtain the exact time evolution by directly performing a matrix multiplication $\ket{\Psi(t)} = U(t) \ket{\Psi_0}$. In the spin basis, the flavor content of an individual neutrino is obtained from the expectation value of the Pauli matrix $ \sigma^{(z)}_i = Z_i $,
\begin{equation}
\expval {Z_i(t)} = \expval{Z_i}{\Psi(t)}\;,
\end{equation}
where the tensor products are implicit for the other particles. In a similar way, the flavor inversion probability $P_i(t)$ can be expressed as
\begin{equation}
P_i(t) = \frac{| \expval{Z_i(0)} - \expval{Z_i(t)} |}{2}\,.
\end{equation}
We display the exact evolution of both quantities in Fig.~\ref{Th_evo}: The top panel shows results for $\expval {Z_i(t)}$ while the bottom panel displays the inversion probabilities. As expected from the exchange symmetry in Eq.~\eqref{Eq_symm_exchange} and the asymmetric choice of initial state $\ket{\Psi_0}$, the flavor evolution of neutrinos $\nu_0$ and $\nu_1$ is the mirror image of neutrinos $\nu_3$ and $\nu_2$ (respectively). This is reflected in the equivalence of inversion probabilities for these neutrinos (bottom panel of Fig.~\ref{Th_evo}). Because of the presence of this symmetry, in the rest of this work we show results for inversion probabilities only.

\section{Implementation of the time evolution operator}
\label{Sec3}
In order to carry out a quantum simulation, one always needs two ingredients: (1) a state encoding map and (2) a way to map operators into quantum gates.
Because of the two-flavor approximation presented in Sec.~\ref{Sec2}, the flavor state of neutrinos can be directly encoded into a qubit according to the map in Eq.~\eqref{encoding}.
In the case of digital quantum simulations, the operator $ U(t) $ must then be decomposed into a sequence of quantum gates from a fixed set.
In this way the initial state encoding the flavor state $ \ket {\Psi_0} $ is evolved under a sequence of unitary transformations, overall implementing the real-time propagator. A projective measurement of the final state of the qubits eventually allows us to extract flavor observables from the simulation.
In the computational basis, the operator $ U (t) $ is represented by a $ 2 ^ N \times 2 ^ N $ unitary matrix which must be decomposed into the gate set provided by the machine, usually composed by single- or two-qubit elementary gates. One way to decompose this operator is to explicitly exploit the actual interaction of the physical system which occurs in pairs and which can therefore be implemented by considering only a pair of qubits at a time. 
The approach followed in Ref.~\cite{hall2021simulation} uses the exact pair propagator. This can be obtained in two steps: First, one symmetrizes the one-body term and expresses the total Hamiltonian as a sum of two-body terms, namely,
\begin{equation}
    H = \sum_{i<j}^{N-1} \bigg( \boldsymbol{b} \cdot \frac{(\boldsymbol{\sigma}_i + \boldsymbol{\sigma}_j)}{N-1} + J_{ij} \boldsymbol{\sigma}_i \cdot \boldsymbol{\sigma}_j \bigg) := \sum_{i<j}^{N-1} H_{ij}\;.
    \label{H_symmetrized}
\end{equation}
Then, the total propagator can be approximated by the product of pair propagators:
\begin{equation}
    U(t)  \approx \prod_{i<j}^{N-1} e^{-iH_{ij}t}\,.
    \label{approx_pair}
\end{equation}
The approximation introduces an error of order $ \mathcal {O} (t ^ 2) $ due to the noncommutativity of the symmetrized two-body terms, namely, $\comm{H_{ij}}{H_{ik}} \not = 0$.
A first improvement of this implementation can be made by considering that the entire one-body and two-body terms commute, namely, $ \comm {H^{(1)}} {H^{(2)}} = 0 $, and therefore their separation does not introduce any errors:
\begin{equation}
\begin{split}
    U(t) = e^{-iH^{(2)}t} e^{-iH^{(1)}t} 
    := U_2(t) U_1(t) \,.
    \label{U1U2}
\end{split}
\end{equation}
Subsequently, the two-body term alone can be efficiently implemented as a pair decomposition:
\begin{equation}
    U_2(t) \approx \widetilde{U}_2(t) = \prod_{i<j}^{N-1} e^{-ih_{ij}t} := \prod_{i<j}^{N-1} u_{ij}(t)\,.
    \label{U2}
\end{equation}
The implementation in Eq.~\eqref{approx_pair} might lead to an increase in the error due to the lack of commutativity between individual one- and two-body contributions and can cause an explicit breaking of the symmetry under particle exchange.
However, the approach can still be useful in the case of a particle-dependent external field, in which $ \boldsymbol {b} \to \boldsymbol {b}_i $, as it occurs when different neutrinos have different energies $E\to E_i$ (a necessary ingredient to observe spectral splits~\cite{duan2010review,Patwardhan2021}).

\subsection{Qubit connectivity and pair ordering}
\begin{figure}
    \centering
    \includegraphics[width = 5cm]{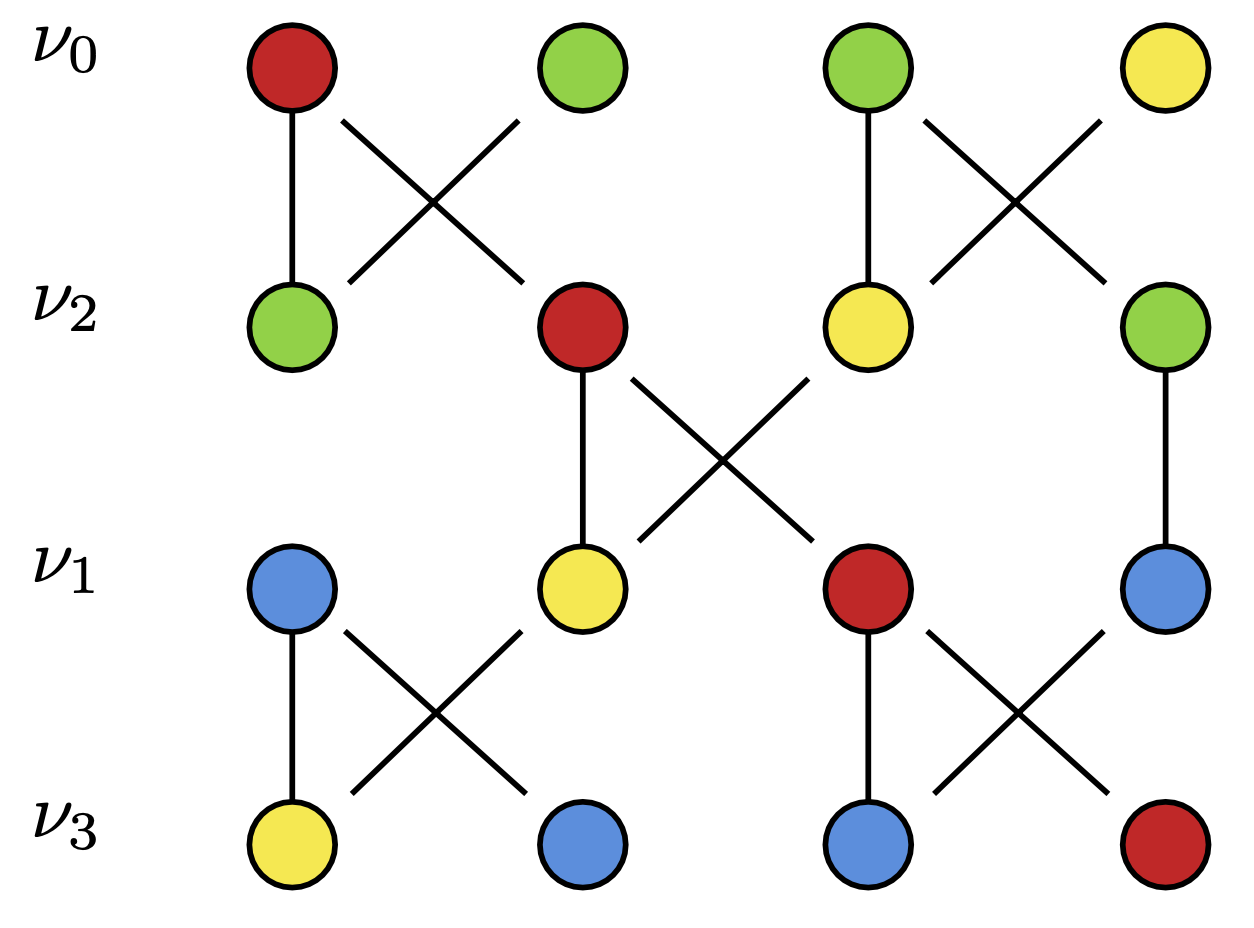}
    \caption{SWAP network for $N = 4$ qubits that implements the two-body propagator $U_2(t)$ using a chain of linearly connected qubits. Vertical lines represent the qubit interaction and crosses the SWAP operation. The pair ordering influences the total error; here, we use the map $(q_0,q_1,q_2,q_3) = (\nu_0, \nu_2, \nu_1, \nu_3)$.}
\label{swap_network}
\end{figure}

\begin{figure}
    \centering
    \includegraphics[width=4cm]{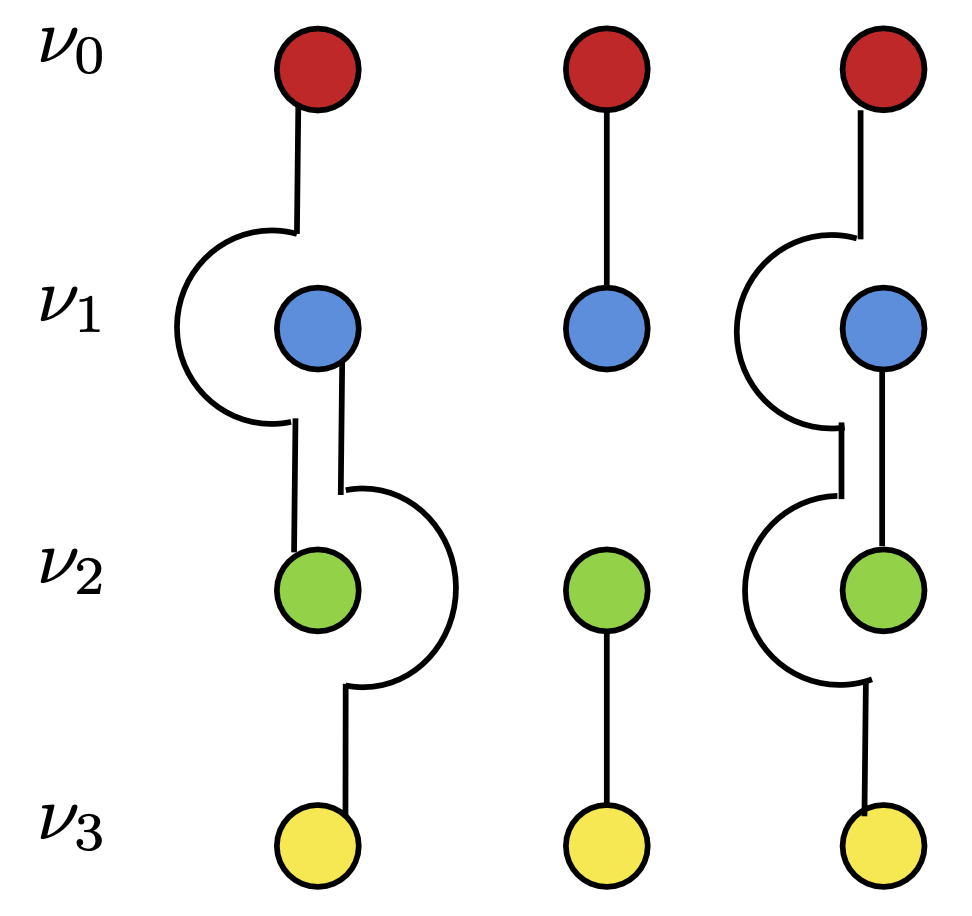}
    \caption{Implementation of the two-body propagator $U_2(t)$ on a qubit system with all-to-all connectivity. This scheme can implement the optimal ordering of pairs to minimize the decomposition error. For our system this corresponds to the qubit to neutrino map $(q_0,q_1,q_2,q_3) = (\nu_0,\nu_1,\nu_2,\nu_3)$.}
    \label{best_network}
\end{figure}

The Hamiltonian in Eq.~\eqref{H} contains an all-to-all interaction term. Using the implementation of $U_2(t)$ from Eq.~\eqref{U2}, we have to make all qubits interact with all the others at least once during the simulation. 
This fact implies that in a quantum computation the circuit implementing the sequence of $U_2(t)$ operators for each particle pair must be adapted to the particular topology of the 
specific quantum device employed for the simulation.
As shown in Ref.~\cite{hall2021simulation} 
it is possible to construct $U_2(t)$ with only linear connectivity and with a gate depth of $N$ using a SWAP network (SN)$\textrm{---}$the same scheme was later adopted for tensor-network simulations in Ref.~\cite{RoggeroMPS2021}. The algorithm consists in 
applying the $u_{ij}$ propagator to a qubit pair followed by a SWAP gate which exchanges the qubit state. The new unitary $w_{ij}$ is thus
\begin{equation}
\label{eq:swap_uij}
w_{ij} = \text{SWAP} \times u_{ij}\;, 
\end{equation}
where in the computational basis the SWAP unitary is
\begin{equation}
\text{SWAP} = \begin{pmatrix}
1&0&0&0\\
0&0&1&0\\
0&1&0&0\\
0&0&0&1\\
\end{pmatrix}\;.
\end{equation}
For example, the SWAP network required for the case of $ N = 4 $ is depicted in Fig.~\ref{swap_network}.
It should be pointed out that 
the error introduced by the approximation in Eq.~\eqref{U2} also depends on the order in which the pairs interact. 
This is due to the dependence of the error in Eq.~\eqref{U2} on sums of the commutators $ \comm {h_{ij}} {h_{kl}} $ taken with a given order (see Appendix~\ref{AppendixA} for additional details). 
It is then worth looking for an optimal ordering
allowing us to maximize the cancellations between the commutators minimizing the decomposition error.
Note that, once the four-layer network structure shown in Fig.~\ref{swap_network} is fixed, the ordering can be chosen by varying the initial encoding of each neutrino into the qubits.
The structure of the SWAP network, dictated by the available qubit topology, imposes a constraint on the possible ordering of the pair propagators. In principle, any ordering could be achieved by adding additional SWAP gates or additional layers but at the cost of  
increasing the complexity in terms of depth and number of two-qubit gates of the scheme. 
For our Hamiltonian, and in the case of $ N = 4 $, we find that the best interaction order would be the one described by the network in Fig.~\ref{best_network}. 
Such ordering cannot be expressed using a SWAP network with four layers (as shown in Fig.~\ref{swap_network}) if restricted to the use of linear qubit connectivity. In fact, one can easily show that at least five layers would be needed.
With all-to-all connectivity, however, this algorithm can be implemented using only three layers, as shown in Fig.~\ref{best_network}. This is due to the fact that each layer is full, in the sense that the maximum number of possible operations at the same time is performed. 
Determining the optimal ordering for large systems is not feasible, in general, as this would require a superexponential cost in the system size $N$.
For large systems a randomization procedure for the order could prove valuable to control the error~\cite{Childs2019fasterquantum,Chen2021}. 

\begin{figure}
    \centering
    \includegraphics[width = 8.5cm]{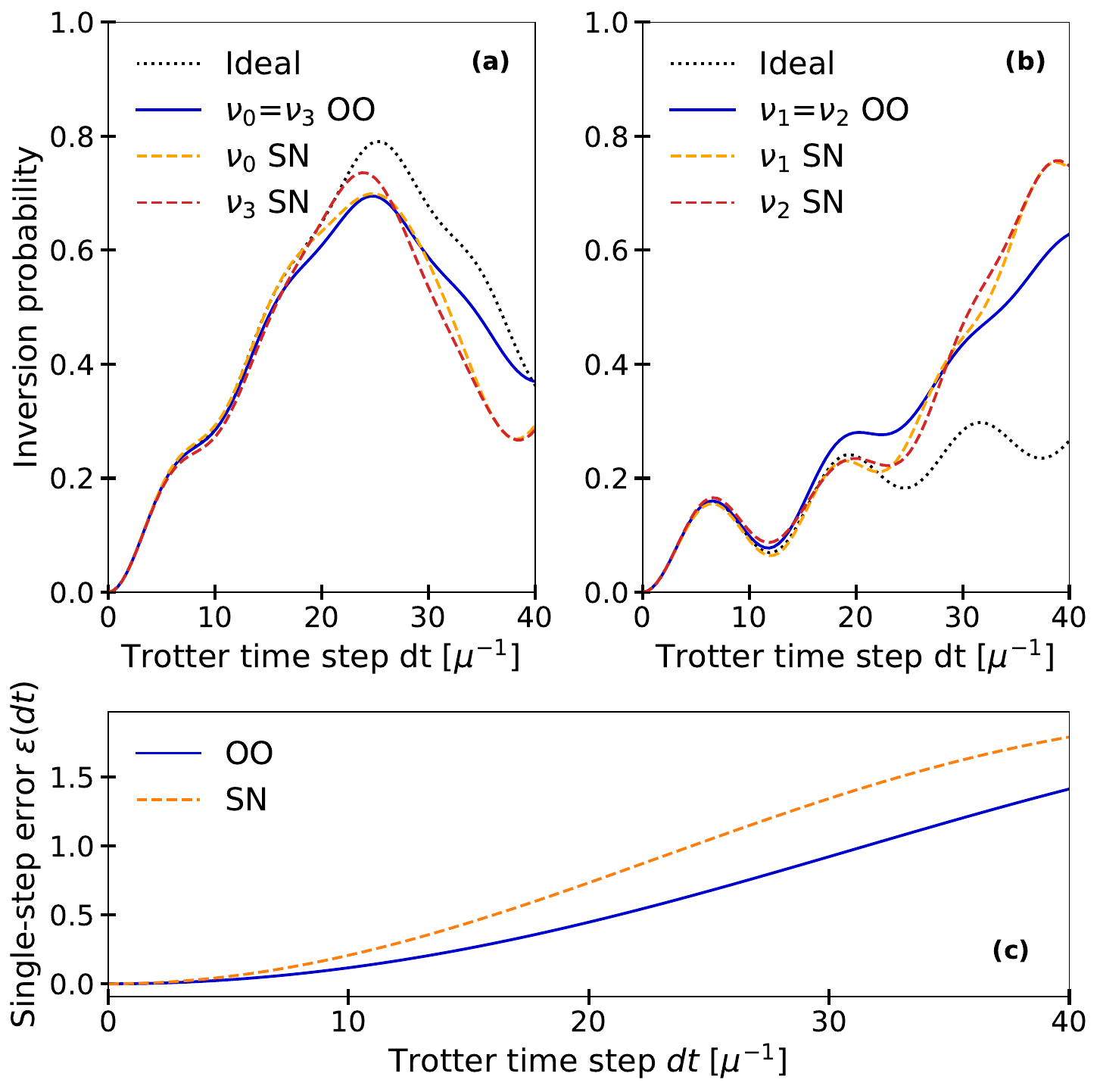}
    \caption{Top panels: inversion probability after a single Trotter step for different values of the time steps $dt$. Panel (a) shows the evolution for neutrinos $\nu_0$ and $\nu_3$, and panel (b) for $\nu_1$ and $\nu_2$. The dotted black curve is the exact evolution, and the solid blue line is the one obtained by applying the implementation proposed in this work [Eqs.~\eqref{U1U2} and \eqref{U2}) together with the scheme in Fig.~\ref{best_network}. The dashed orange and red lines are the evolution using the implementation proposed in Ref.~\cite{hall2021simulation} and described by Eq.~\eqref{approx_pair}, and the scheme from Fig.~\ref{swap_network}. Panel (c) shows the error in spectral norm between the exact propagator and the two Trotter decompositions.
    }
    \label{Single_Trotter_th}
\end{figure}

We analyze the effect of different orderings on a single time step in Fig.~\ref{Single_Trotter_th}. A similar study, made for a different Hamiltonian, about the Trotter error dependence on the ordering can be found in Ref. \cite{nguyen2022digital}. The two top panels show the evolution of the inversion probability as a function of the time step $ dt \in [0,40]\, \mu^{-1} $ for the same initial state as in Fig.~\ref{Th_evo} and for different implementation of the propagator: The dotted curve is the exact evolution, and the solid blue line is the one obtained by applying the implementation proposed in this work [Eqs.~\eqref{U1U2} and \eqref{U2}] together with the optimal ordering (OO) from Fig.~\ref{best_network}. The dashed orange and red lines are the evolution using the implementation proposed in Ref.~\cite{hall2021simulation} and described by Eq.~\eqref{approx_pair}, and the SN scheme from Fig.~\ref{swap_network}. Panel (a) shows results for neutrinos $\nu_0$ and $\nu_3$ while panel (b) shows those for $\nu_1$ and $\nu_2$.
The results highlight the preservation of the exchange symmetry from Eq.~\eqref{Eq_symm_exchange} of the scheme introduced in this work and afforded by the separation between the one- and two-body contributions in Eq.~\eqref{U1U2}. 
Note that this property, observed for $N=4$, does not hold for general system sizes. For instance, we were not able to find an ordering choice preserving this property while keeping the minimum number of layers for the case of $N=8$. Furthermore, the breaking of exchange symmetry in the $N=4$ system for the (SN) results shown in Fig.~\ref{Single_Trotter_th} is to be ascribed to the Trotter breakup from Eq.~\eqref{approx_pair}, which for the (SN) ordering generates an explicit symmetry breaking.

Panel (c) of Fig.~\ref{Single_Trotter_th} shows the error, for a single Trotter step, of the two approximations for the time evolution operator as a function of the time step $ dt $.
We calculate the error using the spectral norm (i.e., the maximum singular value of the matrix)
\begin{equation}
    \varepsilon(dt) = \norm{ \widetilde{U}_2(dt) - U_2(dt)}\,.
    \label{Norm_error}
\end{equation}
The results displayed in the top panels show that the error in the inversion probability is lower for the (SN) approximation when $dt < 30 \, \mu^{-1}$. This is mostly an effect of choosing a particular initial state $\ket{\Psi_0}$ and this specific observable. For general initial states and observables, the error displayed in panel (c) shows that the (OO) approximation has indeed the smallest worst-case error for all time steps. 
\begin{figure}
    \centering
    \includegraphics[width = 8cm]{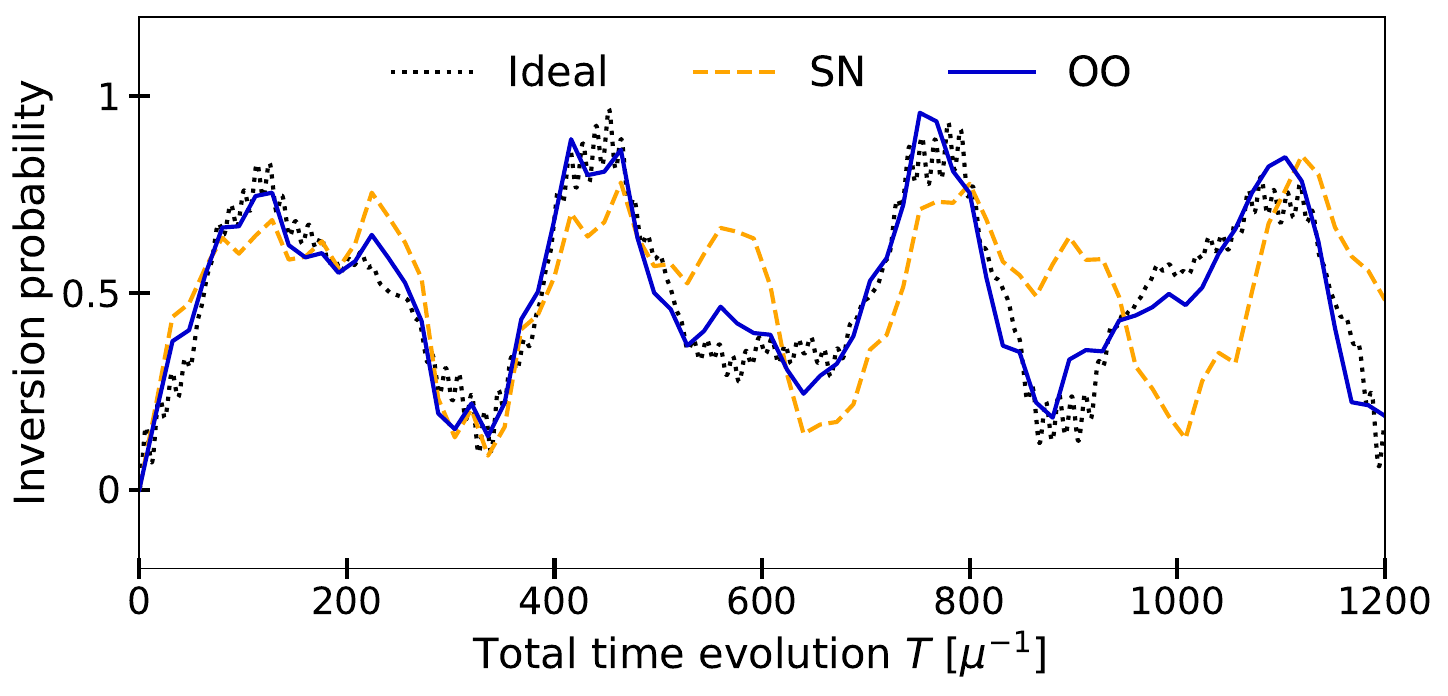}
    \caption{Time evolution of the inversion probability for neutrino $\nu_1$ using a time step $dt = 16 \mu^{-1}$. The dotted black curve is the exact evolution, the orange dashed curve is the Trotter decomposition with the best possible SN while the solid blue curve uses the OO achievable using the full connectivity. 
    }
    \label{long_evolution}
\end{figure}
The main advantage of using the optimal order is evident when the goal of the simulation is to describe the evolution of the system for a long total time $T$ in which the accumulation of the error is dominant. In Fig.~\ref{long_evolution} we plot the time evolution of the inversion probability for the neutrino $ \nu_1 $ (set in the electron flavor $\ket{0}$ at the beginning) for a long total time $ T = 1200\, \mu^{-1}$ using a time step $ dt = 16 \, \mu^{-1}$. For each time, we sequentially apply the approximate propagator
\begin{equation}
    \ket{\Psi(k dt)} = \widetilde{U}_2(dt)^{k} U_1(dt)^k \ket{\Psi_0}\,,
\end{equation}
where $\widetilde{U}_2(dt)$ is defined in Eq.~\eqref{U2}. We employ two different orderings: (1) the best possible ordering that is achievable with a SN with linear connectivity as presented in Fig.~\ref{swap_network} above, and denoted by the dashed orange curve in Fig.~\ref{long_evolution}, and (2) the optimal ordering, achievable by allowing all-to-all connectivity, shown in Fig.~\ref{best_network}, presented as the solid blue curve in Fig.~\ref{long_evolution}. As we can see, the results obtained using OO are much more stable than with SN and allow us to reach long evolution times even with large time steps.

\section{Optimized quantum circuit}
\label{Sec5}
In order to implement the unitary propagator, we need to decompose it as a sequence of elementary gates from the universal gate set used by the quantum machine. The one-body part is trivial because it is the tensor product of the same single-qubit gate applied to each qubit:
\begin{equation}
\label{eq:onebodyu}
    U_1(dt) = \bigotimes_{i=0}^{N-1} \exp\left(-i \boldsymbol{b} \cdot \boldsymbol{\sigma}_i dt\right) \;.
\end{equation}
The two-body part of the propagator, as approximated in Eq. \eqref{U2}, is the product of pair terms of the form
\begin{equation}
\label{eq:su2_uij}
    u_{ij}(dt) = e^{-idt J_{ij}(X \otimes X + Y \otimes Y + Z \otimes Z)}\,.
\end{equation}
In the case of full-qubit connectivity we do not need to add the SWAP gate after the pair interaction [see Eq.~\eqref{eq:swap_uij}] and we can directly implement the $u_{ij}(dt)$ operator. This results in a decomposition with a smaller number of single-qubit gates.
In fact using the result in Ref.~\cite{vatan2004optimal}, the optimal CNOT-based decomposition for the $SU(2)$ invariant unitary operator $u_{ij}$ in Eq.~\eqref{eq:su2_uij} can be written as 
\begin{equation}
    \Qcircuit @C=0.5em @R=1em{
    & \qw & \targ{}& \gate{R_z(\phi)}  & \ctrl{1} & \qw & \targ{} & \gate{R_z(\frac{\pi}{2})} & \qw \\
    & \gate{R_z(-\frac{\pi}{2})} & \ctrl{-1} & \gate{R_y(-\phi)} & \targ{} & \gate{R_y(\phi)} & \ctrl{-1} & \qw & \qw 
    }
    \label{CNOT_circuit}
\end{equation}
where we introduced the angle $\phi=-(2dt J_{ij}+\pi/2)$.

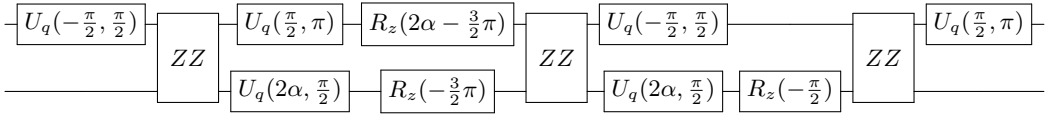
\begin{figure*}
\centering
\begin{equation*}
\Qcircuit @C=0.5em @R=1em{
& \gate{U_q(-\frac{\pi}{2},\frac{\pi}{2})} & \multigate{1}{ZZ} & \gate{U_q(\frac{\pi}{2},\pi)} & \gate{R_z(2\alpha - \frac{3}{2}\pi)} & \multigate{1}{ZZ} & \gate{U_q(-\frac{\pi}{2}, \frac{\pi}{2})} & \qw & \multigate{1}{ZZ} & \gate{U_q(\frac{\pi}{2},\pi)} & \qw \\
& \qw & \ghost{ZZ} & \gate{U_q(2\alpha, \frac{\pi}{2})} & \gate{R_z(-\frac{3}{2}\pi)} & \ghost{ZZ} & \gate{U_q(2 \alpha, \frac{\pi}{2})} & \gate{R_z(-\frac{\pi}{2})} & \ghost{ZZ}  & \qw  & \qw
}
\end{equation*}
\caption{Decomposition of the two-body propagator $u_{ij}(dt)$ in Eq.~\eqref{eq:su2_uij} using native gates and setting $\alpha=-dtJ_{ij}$.}
\label{Opt_circ}
\end{figure*}
When performing a quantum simulation, it is always advisable to write the circuit using the physical gates actually implemented by the machine and optimize it. This procedure allows us to reduce the number of gates necessary for the simulation.
In this work we simulate the real-time evolution of $N=4$ and $N=8$ neutrino systems using the QSM H1-2 trapped-ion device, similar to the machine described in Ref.~\cite{Pino_2021}. Like other trapped-ion-based quantum devices, it offers the advantage of a full connectivity between the qubits, high fidelity: a single-qubit error of $\sim 10^{-4}$ and a two-qubit gate error of $\sim 10^{-3}$, and the possibility to perform parallel operations between different qubits in different zones simultaneously.
The native single-qubit gates used are as follows:
\begin{equation}
    R_z(\lambda) = \begin{pmatrix}
    e^{-i\lambda/2} & 0 \\ 0 & e^{i\lambda/2}
    \end{pmatrix}
\end{equation}
and:
\begin{equation}
    U_q(\theta, \varphi) = \begin{pmatrix}
    \cos{\theta/2} & -i e^{-i\varphi} \sin{\theta/2}\\
    -ie^{i \varphi} \sin{\theta/2} & \cos{\theta/2}
    \end{pmatrix}\,.
\end{equation}
Currently, only two possible values $\theta=\pi,\pi/2$ are natively available on the device. This means that up to four physical gates are required to compile a general $SU(2)$ unitary.
The two-qubit gate is
\begin{equation}
    ZZ = \exp\left(-i \frac{\pi}{4} Z \otimes Z\right) = \begin{pmatrix}
    1 & 0 & 0 & 0 \\
    0 & i & 0 & 0 \\
    0 & 0 & i & 0 \\
    0 & 0 & 0 & 1
    \end{pmatrix}\,.
\end{equation}
In the circuit from Eq.~\eqref{CNOT_circuit}, we can replace the following $ZZ$-based implementation of the CNOT gate:
\begin{equation}
\Qcircuit @C=0.5em @R=0.7em {
    & \ctrl{1} & \qw  & \raisebox{-2.5em}{=}  & & \qw & \multigate{1}{ZZ} & \qw & \gate{R_z(-\frac{\pi}{2})}  & \qw\\
    & \targ & \qw & & & \gate{U_q(-\frac{\pi}{2},\frac{\pi}{2})} & \ghost{ZZ} & \gate{U_q(\frac{\pi}{2}, \pi)} & \gate{R_z(-\frac{\pi}{2})} & \qw .
}   
\end{equation}
Then, using (1) commutation properties, (2) all possible simplifications including cancellation and gate mergers, and (3) translating all single-qubit gates into native gates, we obtain the circuit displayed in Fig.~\ref{Opt_circ} that implements the two-body part of the evolution for one pair of neutrinos using only native gates. 
The angle parameter $\alpha$ is defined, for every qubit pair, as $\alpha=-dtJ_{ij}$.

\begin{figure}[b]
    \centering
    \includegraphics[width = 8.5cm]{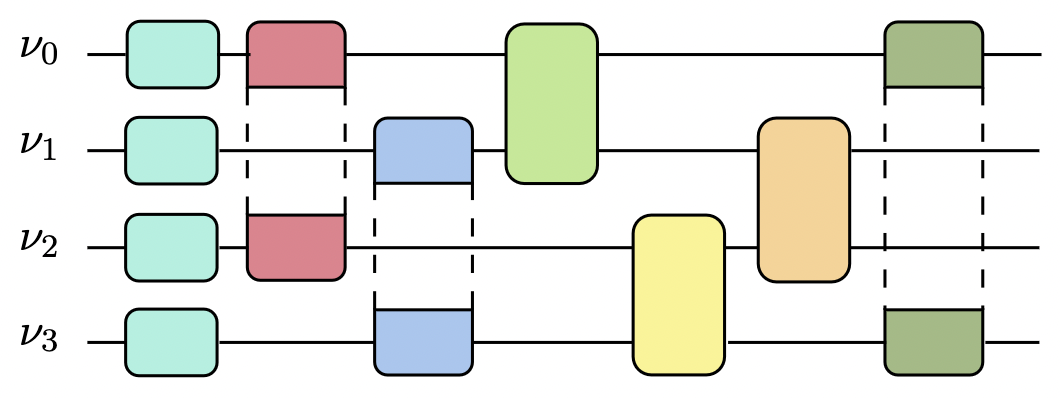}
    \caption{Scheme for implementing a single Trotter step. We first apply the single-qubit gates corresponding to the one-body propagator $U_1(dt)$ and then the pair propagator using the optimal ordering implementing $U_2(dt)$.}
    \label{circuit_total}
\end{figure}

A single Trotter step is implemented by first applying the one-body part \textcolor{blue}{in} Eq.~\eqref{eq:onebodyu}, made up of single-qubit gates, followed by the optimal ordered sequence of pair interactions. A scheme for the circuit of a complete single Trotter step is presented in Fig.~\ref{circuit_total} where each two-qubit gate is implemented by the circuit in Fig.~\ref{Opt_circ}. 
The evolution for a total time $T$ using a time step $dt$ is implemented by sequentially applying the scheme in Fig.~\ref{circuit_total} for $r = T/dt$ times and moving to full one-body part to the beginning.
In these cases the overall gate count for the circuit can be further optimized using an inverted interaction order in alternate steps [this means performing the last interactions $(1,2)+(0,3)$ at the beginning, which is also an equally optimal ordering] resulting in a reduction of $3N/2$ $ZZ$ gates for each step. This alternating scheme can also reduce the overall approximation error since it then becomes equivalent to a second order Trotter step with time step $2dt$ (see Sec.~\ref{Sec4} for additional details).

\section{Results of Quantum Simulations}
\label{Sec6}
In this section we report the results obtained from the simulations carried out on the Quantinuum trapped-ion device.
We perform two distinct simulation: First, we approximate the full time evolution with a single Trotter step with different values of $dt$, as done in Ref.~\cite{hall2021simulation}. Second, we perform a multistep simulation fixing $dt=4 \, \mu^{-1}$. 
The single-step results are obtained using both a system with $N=4$ neutrinos$\textrm{---}$equivalent to the one studied in Ref.~\cite{hall2021simulation} on a superconducting device and in Ref.~\cite{illa2022basic} on a quantum annealer$\textrm{---}$and a larger system with $N=8$. Similarly to this work, the simulation on superconducting circuits required a number of qubits equal to $N$, while, using the strategy described in Ref.~\cite{illa2022basic}, in order to map the problem into a quantum annealer, we would need at least $2^{N+2}$ physical qubits in the ideal case of all-to-all connectivity (corresponding to $64$ qubits for $N=4$ and $1024$ qubits for $N=8$). However, due to the limited connectivity in these devices, a large number of auxiliary qubits need to be employed. With the Pegasus topology implemented in the D-Wave Advantage system, the $N=4$ simulation presented in Ref.~\cite{illa2022basic} required $\approx2000$ physical qubits; by extrapolating the resource requirements estimated in Fig.~9 there, a simulation with $N=8$ neutrinos will require $\approx10^5$ physical qubits and is likely out of reach in the near future. 

For each simulation, the circuit is repeated  $M=200$ times in order to collect statistics for the measurement outcomes. The results of the simulation are then analyzed by calculating 
statistical confidence intervals using the Bayesian approach already employed in Ref.~\cite{hall2021simulation}. 

\subsection{Single Trotter step propagation}
The initial flavor states for $N=4$ and $N=8$ are chosen to contain a mixture of both $e$ and $x$ flavors as
\begin{equation}
\rvert\Psi_0^{(4)}\rangle = \ket{0011}\;,\quad\quad    \rvert\Psi_0^{(8)}\rangle = \ket{00001111}\;,
\end{equation}
respectively, and are prepared using single-qubit gates $U_q$. We then apply the Trotter step according to the diagram displayed in Fig.~\ref{circuit_total} (and its generalization to the $N=8$ case) using the gate decomposition in Fig.~\ref{Opt_circ}. This requires a number of two-qubit $ZZ$ gates equal to $18$ (for $N=4$) and $84$ (for $N=8$). Since the trapped-ion device has more than four qubits at our disposal, in order to minimize the execution cost of the $N=4$ experiments, we always use two sets of four qubits at the same time, each one performing the circuit sequence corresponding to two different values of $dt$. 
Thus, the results presented below are all obtained from experiments using eight qubits with a two-qubit gate depth given by either $9$ or $21$ (for the $N=4$ and $N=8$ simulations, respectively). 
\begin{figure}
    \centering
    \includegraphics[width = 8.5cm]{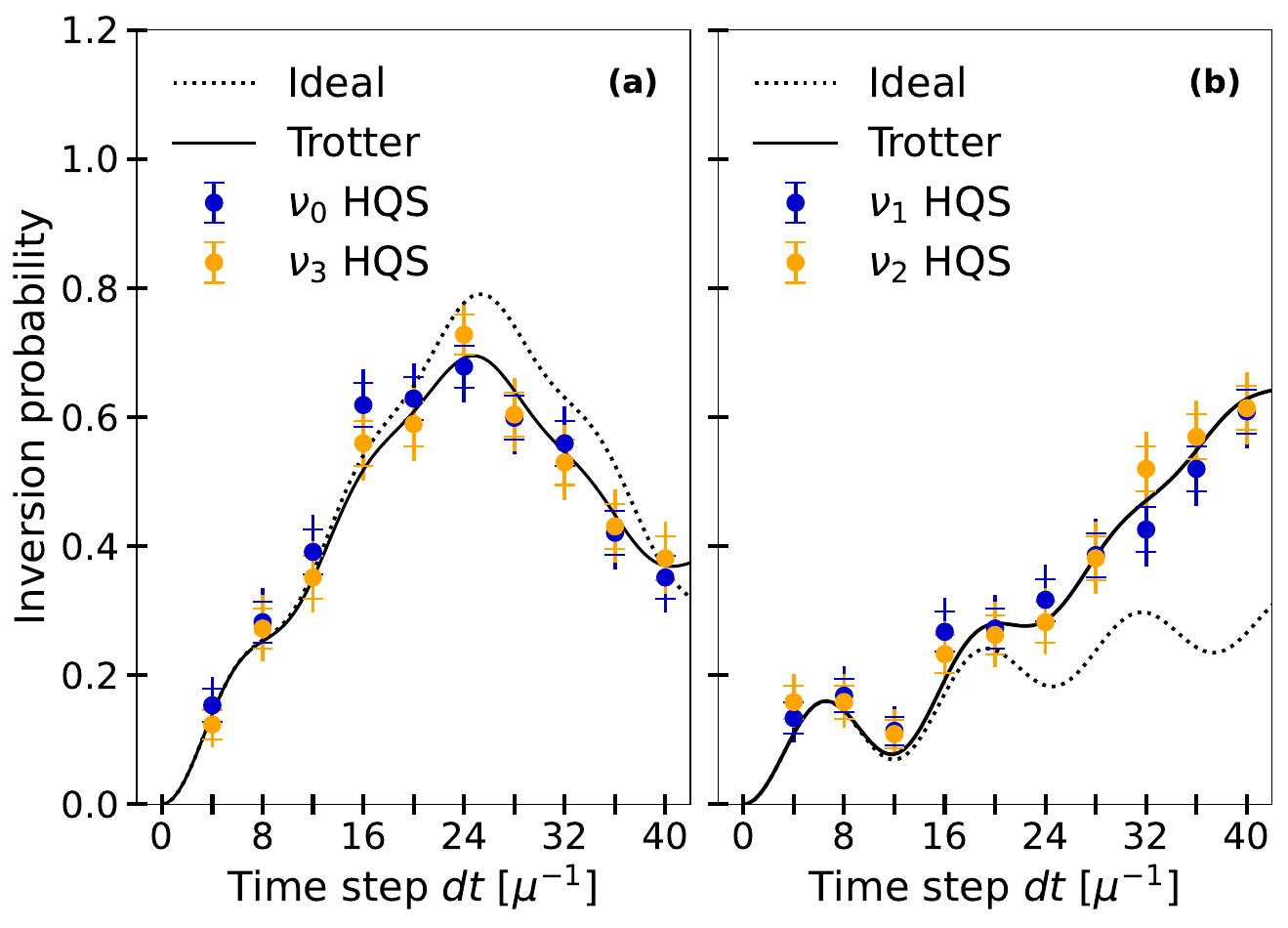}
    \caption{Single Trotter step evolution for the inversion probability starting with the initial state $\rvert\Psi_0^{(4)}\rangle=\ket{0011}$. Panel (a) is for neutrinos $\nu_0$ and $\nu_3$, and panel (b) is for $\nu_1$ and $\nu_2$.
    The dotted line represents the ideal results using the exact propagator, while the solid black line indicates the ideal result obtained using one Trotter step. The results obtained from experiments on QSM H1-2 are represented by data points and error bars with and without caps corresponding to $68\%$ and $90\%$ confidence intervals, respectively.}
    \label{Trotter4}
\end{figure}
\begin{figure}
    \centering
    \includegraphics[width = 8cm]{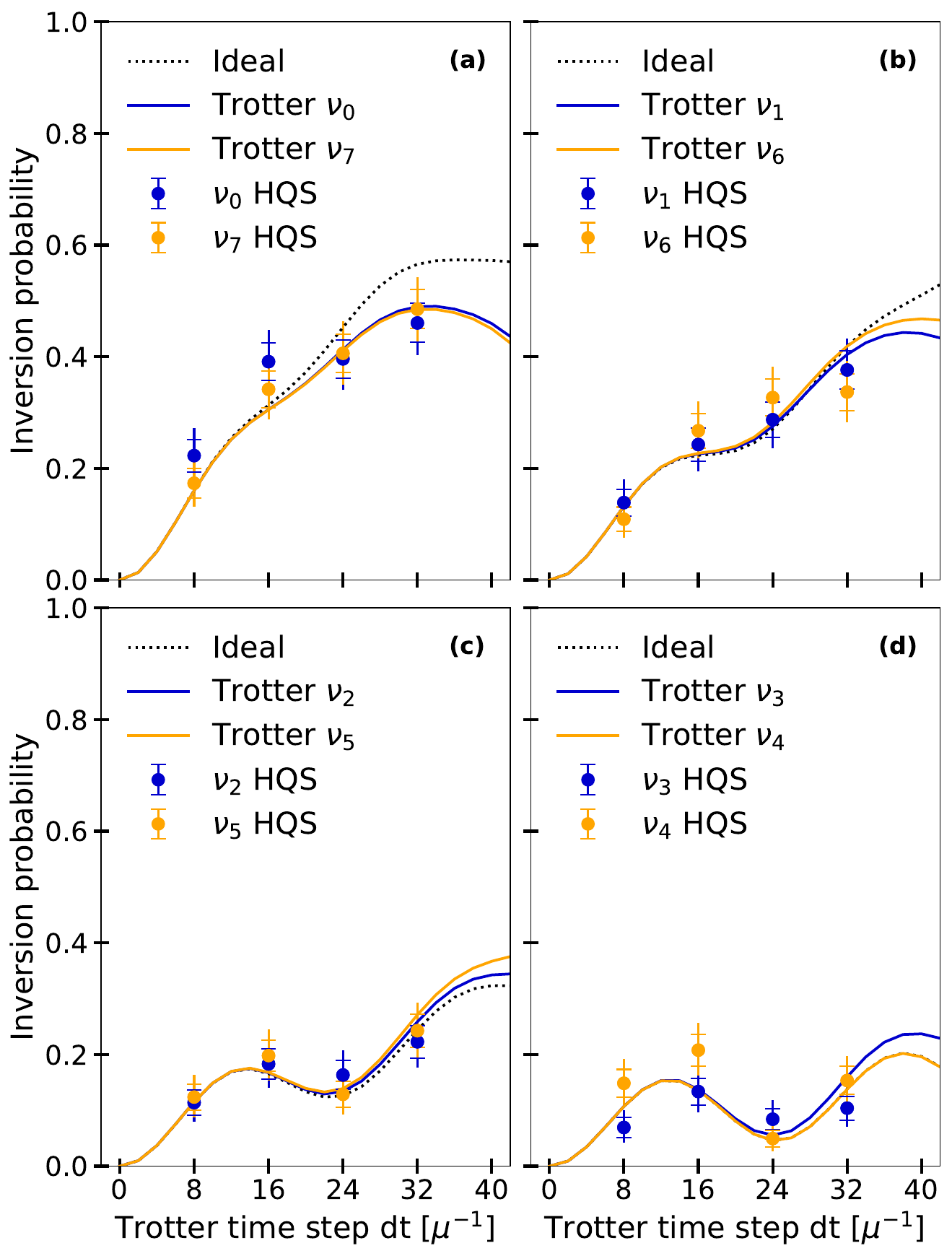}
    \caption{Single Trotter step evolution for the inversion probability starting with the initial state $\rvert\Psi_0^{(8)}\rangle=\ket{00001111}$. Panel (a) shows result for the pair $(\nu_0,\nu_7)$, panel (b) for $(\nu_1,\nu_6)$, panel (c) for $(\nu_2,\nu_5)$, and panel (d) for $(\nu_3,\nu_4)$. Curves and data points follow the same convention used in Fig.~\ref{Trotter4}.}
    \label{8neutrini}
\end{figure}

Figure~\ref{Trotter4} shows the results for the inversion probability obtained for the simulation of a system of $N=4$ neutrinos while Fig.~\ref{8neutrini} shows those for a system of $N=8$ neutrinos. 
In both figures the pairs of neutrinos related by the exchange symmetry from Eq.~\eqref{Eq_symm_exchange} are represented in the same panel: neutrinos $\nu_0$ and $\nu_7$ in panel (a), $\nu_1$ and $\nu_6$ in panel (b), $\nu_2$ and $\nu_5$ in panel (c), and $\nu_3$ and $\nu_4$ in panel (d). 
In the limit of negligible machine errors, the implementation of the propagator proposed in this work guarantees perfect symmetry 
under particle exchange in the case of $N=4$, and the results obtained on the real device almost always respect this symmetry within a confidence interval of $ 68 \% $ and always in the interval of $ 90 \% $.
Because of our choice of using eight qubits to perform two simulations for $N=4$ in parallel, these errors might, in principle, be affected by the cross-talk between the two simulations. 
However, by analyzing the results of three parallel executions over $12$ qubits using time steps $dt = 16,\,24,\,32\,\mu^{-1}$, we found them to be compatible with the results shown in Fig.~\ref{Trotter4}, indicating that cross-talk effects are minimal.

For $N=8$ neutrinos our implementation of the propagator respects the symmetry of particle exchange up to $dt \approx 24 \, \mu^{-1}$ as can be seen by the theoretical results shown as solid lines in Fig.~\ref{8neutrini}. The real data respect the same symmetry within the $90 \%$ confidence interval. The results obtained for the evolution of a single step are very promising and much more compatible with the theoretical ones than those obtained in Ref.~\cite{hall2021simulation}.

In order to compare the results obtained in this work using a trapped-ion device and with optimal ordering OO to the previous results obtained using a swap network SN on the IBMQ Vigo superconducting device~\cite{hall2021simulation}, in Table~\ref{chi} we present the values for $\chi^2$ of the inversion probabilities of each neutrino for the $N=4$ simulation. 
The quality of the results is assessed by measuring the distance from the theoretical prediction of the inversion probability of neutrino $i$ at time $t$, $P^{(th)}_i (t)$, of the computed results $P_i(t)$ by means of the following function: 
\begin{equation}
\chi_i^2 = \frac{1}{10} \sum_ {k = 1} ^ {10} \frac{\left(P_i (kdt) - P^{(th)}_i (kdt)\right) ^ 2} {\delta P_i (kdt) ^ 2 }\;,
\end{equation}
where $\delta P_i(t)$ is the estimated variance (taken to coincide with the $68\%$ confidence interval) and $10$ is the number of simulated points, for each neutrino, used to calculate $\chi^2$. 
The \textcolor{blue}{\sout{last} third} row shows results from the superconducting IBMQ device after error mitigation. In our current simulation we did not attempt to mitigate errors and the report values are the bare results. It can be noticed that there is an important increase of the fidelity in the results presented in this work. This is a combined effect of the higher gate fidelity provided by the trapped-ion device and the reduction in complex one-qubit rotations afforded by the different Trotter decomposition from Eq.~\eqref{U1U2} adopted here.
The number of general $SU(2)$ single-qubit operations in a single step for $N=4$ is in fact reduced from $40$  with the decomposition adopted in Ref.~\cite{hall2021simulation} to $36$ in the present work. More importantly, the $40$ rotations adopted in the previous work have arbitrary angles due to the combination of the pair propagator and the SWAP gate and always require three elementary rotations to be implemented. Exploiting instead the full connectivity and the optimal decomposition in Fig.~\ref{Opt_circ}, about 2/3 of the unitaries consist of rotations of angles which are multiples of $\pi/2$.  This helps reduce the effect of coherent errors in the final results. Finally, the results of the simulations carried out on IBMQ are obtained using a much larger statistical sample (8192 circuit repetitions instead of 200), in order to more directly compare the new results with the ones obtained there; in the last line of Table~\ref{chi}, we also report the estimated $\chi^2$ we would have expected to see if we reduced the statistics of the bare IBMQ results. The same procedure cannot be consistently performed for the mitigated results since such an estimate is also affected by systematic errors (for more details see Refs.~\cite{hall2021simulation,Roggero_nptodg,roggero2020A}). However, the strong effect of an increased gate fidelity is still evident.

\begin{table}[t]
    \centering
    \begin{tabular}{l|c c c c }
    \hline \hline
         & $\nu_0$ & $\nu_1$ & $\nu_2$ & $\nu_3$ \\
        \hline
        OO + QSM H1-2 (bare) & 0.36 & 0.35 & 0.27 & 0.14\\
        SN + IBMQ Vigo (bare) & 424.31 & 527.64 & 545.28 & 502.60\\
        SN + IBMQ Vigo (mit) & 71.35 & 73.64 & 126.38 & 142.72\\
        SN + IBMQ Vigo (bare*) & 10.36 & 12.88 & 13.31 & 12.27\\
        \hline \hline
    \end{tabular}
    \caption{Values of $ \chi ^ 2 $ for each neutrino calculated on the results obtained from the propagation of a single Trotter step. The results denoted IBMQ Vigo are taken from Ref.~\cite{hall2021simulation} while the QSM H1-2 ones are from the present work.}
    \label{chi}
\end{table}

\subsection{Multiple Trotter steps for $N=4$}
In order to reach long simulation times while keeping the error under control, the standard approach is to divide the full interval into time steps which are then approximated using a short-time approximation. For a 
system of $ N = 4 $ neutrinos, initially in the state $ \ket{\Psi_0}$, this can be done through the sequential application of the scheme depicted in Fig.~\ref{circuit_total} for a number of 
time steps $k$, obtaining the final state 
\begin{equation}
\ket {\Psi (kdt)} = \widetilde{U}_2 (dt) ^ k U_1 (dt) ^ k \ket {\Psi_0}\,.
\end{equation}
For the results shown in this section, we used $k\in[1,10]$ with a time step $ dt = 4 \, \mu^{-1} $.
Also in this case, in order to exploit the specific features of the trapped-ion machine, we have carried out the simulations in pairs in which the first four neutrinos are evolved up to a time $ T = kdt $ and the last four up to $ T = (k + 1) dt $. This means that the circuit to simulate the first two times $ T = 1$ and $ T = 2 $ contains 18 $ ZZ $ gates applied to the first four qubits and 36 to the last four. 
We summarize the gate counts needed to simulate the system up to a certain evolution time $ T = kdt $, divided into the number of two-qubit $ZZ$ gates and one-qubit $SU(2)$ unitaries, in Table~\ref{number_gates}. 
For each step we need $6\times3$ two-qubit gates while, exploiting cancellations between neighboring pair propagators, the number of general single-qubit rotations needed scales as $32\times k+4$.
\begin{table}[b]
    \centering
    \begin{tabular}{l|c c c c c c c c c c }
    \hline \hline
        $\#$ of steps $k$ & 1 & 2 & 3 & 4 & 5 & 6 & 7 & 8 & 9 & 10\\
        \hline
        $\#$ of $ZZ$ gates & 18 & 36 & 54 & 72 & 90 & 108 & 126 & 144 & 162 & 180\\
        $\#$ of $SU(2)$ gates & 36 & 68 & 100 & 132 & 164 & 196 & 228 & 260 & 292 & 324 \\        \hline \hline
    \end{tabular}
    \caption{Number of $SU(2)$ gates and $ZZ$ gates to evolve the system to a fixed time $T = k dt$ to produce the results in Fig. \ref{Evolution_real}. }
    \label{number_gates}
\end{table}
\begin{figure}
    \centering
    \includegraphics[width =9cm]{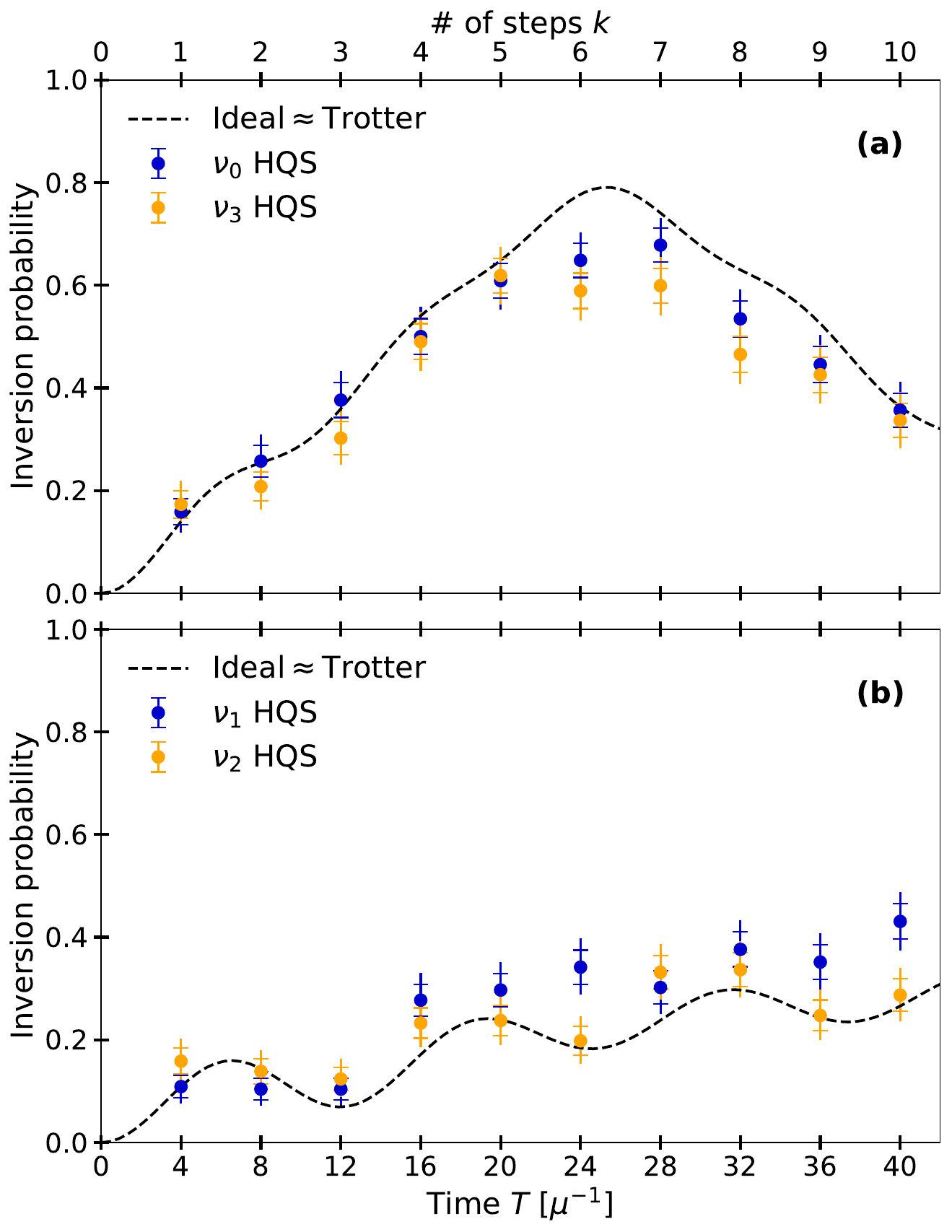}
    \caption{Real-time evolution of a system with $N=4$ neutrinos for time step $dt = 4 \, \mu^{-1}$ and for a total time of $T = kdt$ with $k \in [1,10]$ using the QSM H1-2 trapped-ion device. The top panel shows neutrinos $\nu_0$ and $\nu_3$ and the bottom one $\nu_1$ and $\nu_2$. Dashed black lines are the ideal evolution, which, using this small time step, is almost the same as the Trotter approximated propagation.}
    \label{Evolution_real}
\end{figure}

In this quantum machine, measurements are carried out by default at the end of the circuit. Our simulations include two independent circuits applied on two separate blocks of qubits where the first ends before the second. This means that the first four qubits are subject to idle errors which, in the case of trapped ions, are mainly related to dephasing. One way to improve the simulation is to use a Dynamical Decoupling approach \cite{das2021adapt} consisting in applying to idle qubits a sequence of single-qubit gates that have the overall effect of an identity. 

\section{Error scaling and gate cost}
\label{Sec4}
In this section we analyze the scaling of the Trotter error and the gate cost of the circuit necessary to evolve a system of neutrinos as a function of the number $N$ of neutrinos.
To compare the different decomposition techniques, we look at the scaling corresponding to a description of a system in which we fix the neutrino density $n_{\nu} = N/V$ and increase the dimension of the system.
Additional details on the results presented here can be found in the appendices. 

The implementation in Eq. \eqref{U2} approximates the two-body part of the propagator $U_2(dt) = e^{-iH^{(2)}dt}$ using a standard first order Trotter formula:
\begin{equation}
    \mathcal{L}_1(dt) = \prod_{i<j}^{N-1}e^{-ih_{ij} dt}\;.
    \label{eq:first}
\end{equation}
By leveraging the results in Ref.~\cite{childs2021theory} (see Appendix~\ref{AppendixA} for additional details), we can bound the spectral norm error in this approximation as
\begin{equation}
\varepsilon_1(dt) := \norm{\mathcal{L}_1(dt) - e^{-iH^{(2)}dt}} \leq \mathcal{O}(dt^2\mu^2N)\;.
\end{equation}
Using the union bound to obtain $\varepsilon_1(T)\leq r\varepsilon_1(dt)$ with $dt=T/r$, one also finds that the total number of steps $r_1$ required to evolve for a final time $T$ while keeping the total error below $\epsilon$ scales linearly with system size.
We can then bound the gate cost $\mathcal{C}_1$ of a quantum circuit by implementing the full evolution in terms of the number of  general $SU(4)$ two-qubit gates needed to implement all the steps (which, on the trapped-ion system used for this work, could be decomposed as shown in Fig.~\ref{Opt_circ}). For $N$ neutrinos we have $N(N-1)/2$ interaction terms in each Trotter steps, so the gate cost using a first order product formula scales with
\begin{equation}
\begin{split}
    \mathcal{C}_1 &=\frac{N(N-1)}{2}r_1\leq \mathcal{O}\bigg(\frac{T^2 \mu^2 N^3}{\epsilon}\bigg)\,.
    \label{C1}    
\end{split}
\end{equation}
Since $SU(4)$ transformations can be decomposed with at most three entangling gates~\cite{vatan2004optimal}, as we did for the native gate set available on the QSM H1-2 system in Fig.~\ref{Opt_circ}, the total count for two-qubit gates is given by $3\mathcal{C}_1$. 

\begin{table*}
    \centering
    \begin{tabular}{ l|c c c }
    \hline \hline
        Decomposition type & Single-step error & Number of steps & Circuit complexity\\
        \hline 
        First order Trotter & $\mathcal{O}(dt^2 \mu^2 N)$ & $\mathcal{O}\left(\frac{T^2 \mu^2}{\epsilon} N \right)$ & $\mathcal{O}\left(\frac{T^2 \mu^2}{\epsilon} N^3 \right)$\\
        Second order Trotter & $\mathcal{O}(dt^3 \mu^3 N)$ & $\mathcal{O}\left(\frac{T^{3/2} \mu^{3/2}}{\sqrt{\epsilon}}\sqrt{N} \right)$ & $\mathcal{O} \left(\frac{T^{3/2} \mu^{3/2}}{\sqrt{\epsilon}}N^{5/2} \right)$ \\
        Qubitization & - &$\mathcal{O}$ $\left( T \mu N + \log({1/\epsilon})\right)$ & $\mathcal{O}$ $\left( T \mu N^3 + \log({1/\epsilon}) N^2\right)$  \\
        \hline \hline
    \end{tabular}
    \caption{Asymptotic scaling of the error, needed steps, and number of two-qubit operations to evolve a system until $T$, keeping the error below $\epsilon$ as a function of the number of particles $N$ and for different methods of propagator decomposition.}
    \label{scaling_gates}
\end{table*}

It is possible to obtain a more accurate approximation by using a second order Trotter-Suzuki formula which can be expressed compactly as
\begin{equation}
    \mathcal{L}_2(dt) = \mathcal{L}_1\left(\frac{dt}{2}\right)\mathcal{L}^\dagger_1\left(-\frac{dt}{2}\right)\;.
    \label{eq:second}
\end{equation}
As shown in more detail in Appendix~\ref{AppendixC}, one can also show that in this case the single-step error scales at most linearly with $N$. In particular,
\begin{equation}
\varepsilon_2(dt) \leq \mathcal{O}( dt^3 \mu^3 N)\;.
\end{equation}
In this case, the number of steps needed to guarantee a total error $\epsilon$ for simulation up to $T = rdt$ scales as
\begin{equation}
r_2\leq\mathcal{O}\left(\left(T \mu\right)^{3/2} \sqrt{\frac{N}{\epsilon}}\right)\;.
\end{equation}
Since the number of $SU(4)$ gates is now $N(2N-3)/2$ (assuming, as before, $N$ even) the total gate cost is now
\begin{equation}
    \mathcal{C}_2 = \frac{N(2N-3)}{2}r_2 \leq \mathcal{O} \left( \frac{(T \mu)^{3/2}}{\sqrt{\epsilon}} N^{5/2} \right)\;.
\end{equation}
The scaling with the number of neutrinos $N$ has substantially improved within this scheme. The use of even higher order formulas could allow one to reach an almost optimal scaling $\mathcal{C}(N^{2+\delta})$ for $\delta\ll1$, but with possibly much larger constant prefactors.

Finally, we comment on the prospect of using more modern approaches to simulate the time evolution operator, such as qubitization~\cite{qsp2017,low2019hamiltonian}.
This scheme also approximates the propagator over small time intervals but, contrary to Trotter-Suzuki-based approaches, is able to reach an optimal scaling in both $T$ and the error $\epsilon$ for the number of steps
\begin{equation}
r_{Q} \leq \mathcal{O}\left(T\alpha_{H}+\log\left(\frac{1}{\epsilon}\right)\right)\;.
\end{equation}
Here, $\alpha_H$ is a suitable norm of the Hamiltonian operator which, for the two-body neutrino Hamiltonian, is given by
\begin{equation}
\alpha_H = 3\frac{\mu}{N}\sum_{i<j}^N\left(1-\cos(\theta_{ij})\right)=\mathcal{O}(\mu N)\;.
\end{equation}
For a general angular distribution, the gate cost for each segment scales as the number of different coefficients, and thus the gate cost for the algorithm scales as
\begin{equation}
    \mathcal{C}_Q \leq \mathcal{O} \bigg(  T \mu N^3 + N^2\log(\frac{1}{\epsilon})\bigg) \;.
    \label{Cq}
\end{equation}
For a fixed evolution time $T$ and target error $\epsilon$, the second order Trotter-Suzuki scheme then scales better than a qubitization-based approach. This is not a special property of the neutrino system; it has been noted already in other applications (e.g., simulations of the Schwinger model~\cite{Shaw2020,Rajput21}) and is related to the fact that qubitization does not exploit the commutation properties of the terms that form the Hamiltonian.
We summarize the results on the bounds for both the number of time steps and the circuit (gate) complexity of a simulation with $N$ neutrinos, maximum time $T$, and error tolerance $\epsilon$ in Table~\ref{scaling_gates}.

\begin{figure}[h!]
    \centering
    \includegraphics[width = 8cm]{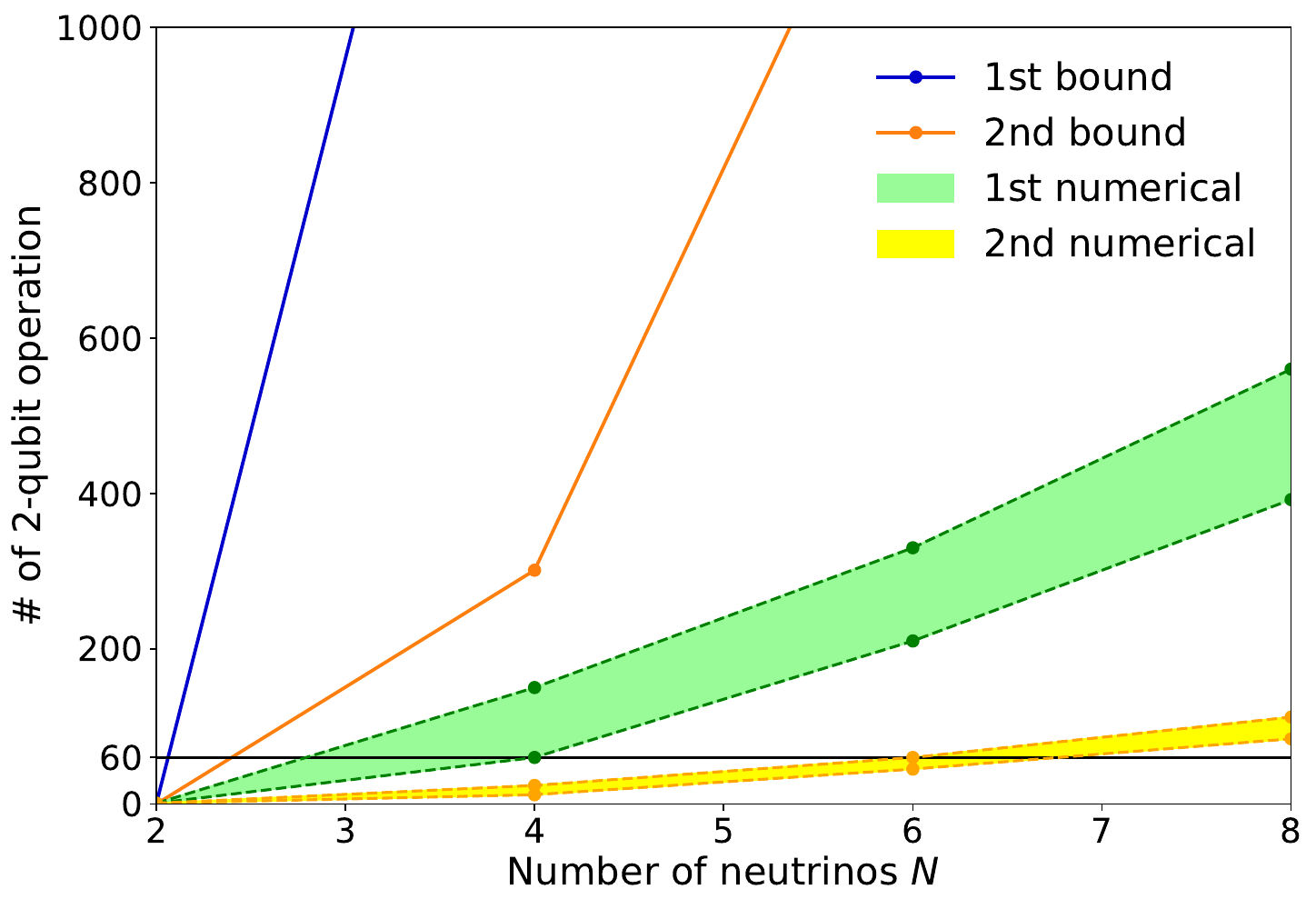}
    \caption{Number of two-qubit operations $\mathcal{C}$, as a function of $N$, needed to evolve the system of a total time $T = 40 \, \mu^{-1}$ with an error $\varepsilon_1(T) \leq \epsilon = 0.15$. The solid blue line corresponds to the first order bound and the green band to the real scaling achievable using different pair ordering. The solid orange line refers to the second order bound instead, and the yellow band is the corresponding real achievable complexity. Using the optimal decomposition, for the system of $N=4$ neutrinos, the number of two-qubit universal operations needed is equal to $10N(N-1)/2 = 60$ and corresponds to $180$ CNOT or $ZZ$ gates, which we actually use to obtain the last point in Fig. \ref{Evolution_real}.}
    \label{mycost}
\end{figure}

The actual implementation cost is several orders of magnitude lower than what is predicted by the theoretical bounds. Moreover, as demonstrated in Sec. \ref{Sec3}, the cost can be further reduced by using a good decomposition of the propagator which guarantees a smaller error and therefore allows us to use a greater value of time step $ dt $. 
For a total evolution time $T=40\, \mu^{-1}$ and target error $\epsilon=0.15$, we show in Fig.~\ref{mycost} the theoretical bounds' gate count $ \mathcal {C}_1 $ and $\mathcal{C}_2$ for the first and second order decompositions as solid blue and orange lines, respectively. We also numerically determine the actual range of gate counts required for this simulation as we vary the order in the Trotter decomposition, which we show in Fig.~\ref{mycost} as the shaded green and yellow bands for first and second formulas respectively. The real complexity is calculated using a linear accumulation error; that is, we search $dt$ such that $r=T/dt$ guarantees
\begin{equation}
    r\norm{\widetilde{U}_{2}(T/r) - U_2(T/r)} \leq \epsilon = 0.15\,,
\end{equation}
where $\widetilde{U}_{2}(T/r) = \mathcal{L}_1(T/r)$ or $\widetilde{U}_{2}(T/r) = \mathcal{L}_2(T/r)$ as defined in Eqs. \eqref{eq:first} and \eqref{eq:second}, respectively.
Using the exact accumulation of error,
\begin{equation}
    \norm{\widetilde{U}_2(T/r)^r - U_2(T)} \leq 0.15
\end{equation}
the number of needed steps could potentially be smaller.

\section{Conclusion}
In this work we
considerably extended the work of  Ref.~\cite{hall2021simulation} in terms of designing 
efficient ways to simulate through a quantum computer the time evolution of a system of several neutrinos interacting with an all-to-all Hamiltonian, demonstrating that the complexity of the algorithm scales in a polynomial way with the number of particles.
In particular, we have shown that the choice of an optimal order in the decomposition can decrease the Trotter error and therefore the number of steps necessary to simulate the evolution of the system for a certain time. This is possible thanks to the use of a machine with full connectivity between the qubits. 
Devices of this type allow us to implement the evolution in circuits with a smaller number of single- and two-qubit gates, potentially decreasing the computational error.
The proposed algorithm was tested on the Quantinuum H1-2 quantum computer to carry out the evolution of a system of $ N =4 $ and $ N = 8 $ neutrinos.
The device employed had very good fidelity gates, and the results obtained, even without error mitigation, have remarkably small errors even in the largest circuits we implemented, containing up to $616$ $SU(2)$ rotations and $342$ two-qubit gates on eight qubits. In the previous study from Ref.~\cite{hall2021simulation}, error mitigation using zero noise extrapolations was instead crucial to obtain acceptable results, even for circuits with a significantly lower number of operations (cf. Table~\ref{chi} here).

Future improvements could be obtained using various error mitigation techniques, such as symmetry protection~\cite{tran2021faster}, virtual distillation~\cite{huggins2021virtual}, and symmetry verification~\cite{bonet2018low}.
Considering the quality of the results obtained in this work, we conclude that it would be possible to both increase the number of neutrinos in the system and to evolve it for a greater total time $T$.
In order to describe more realistic and phenomenologically rich neutrino systems, it will be important to extend the algorithms presented here in order to simulate neutrinos with different energies, which can be modeled with a particle-dependent external field.
Furthermore, another issue to investigate is the description of collective oscillations in the presence of electrons, thus including the matter part in the Hamiltonian.\\

\begin{acknowledgements}
This research used resources of the Oak Ridge Leadership Computing Facility, which is a U.S. Department of Energy Office of Science User Facility supported under Contract No. DE-AC05-00OR22725. F.T. is supported by the Q@TN grant ANuPC-QS. P.L. is supported by the Q@TN grant ML-QForge. This work was prepared in part by LLNL under Contract No. DE-AC52-07NA27344 with support from the Laboratory Directed Research and Development Grant No. 19-DR-005.
\end{acknowledgements}

\appendix
\section{First order Trotter error for the interaction Hamiltonian}
\label{AppendixA}

The 2-body part of the neutrino Hamiltonian in Eq. \eqref{H} of the main text is a sum of $\Gamma = N(N-1)/2$ terms:
\begin{equation}
    H^{(2)} = \sum_{i<j}^N h_{ij} := \sum_{K=1}^{\Gamma} h_K\,,
\end{equation}
where $h_{ij} = J_{ij} \boldsymbol{\sigma}_i \cdot \boldsymbol{\sigma}_j$ and $J_{ij} = \mu(1 - \cos(\theta_{ij}))/N$. Note that we count here the neutrinos from $1$ to $N$ instead of from $0$ to $N-1$.
Using the first order Trotter decomposition we can implement the propagator with a product formula:
\begin{equation}
    U_2(dt) \approx \mathcal{L}_1(dt) = \prod_{i<j}^N e^{-ih_{ij} dt} = \prod_{K=1}^{\Gamma} e^{-ih_K dt}\,.
\end{equation}
Using the result from Proposition 9 of Ref.~\cite{childs2021theory} we bound the first order Trotter error as:
\begin{equation}
\begin{split}
    \varepsilon_1(dt) &= \norm{\mathcal{L}_1(dt) - e^{-iH^{(2)}dt}} \\
    &\leq \frac{dt^2}{2} \sum_{K=1}^{\Gamma}\norm{\sum_{L=K+1}^{\Gamma}  \comm{ h_K}{h_L} }\,,
\end{split}
    \label{Trotter_formula1}
\end{equation}
where in our case $K$ and $L$ correspond to pair indices $K = (i,j)$ and $L = (k,l)$. The sum inside the norm in the expression above can be expressed explicitly as
\begin{equation}
\begin{split}
\sum_{L=K+1}^{\Gamma}  \comm{ h_K}{h_L}&=\delta_{ik} \sum_{l=j+1}^N \comm{h_{ij}}{h_{kl}}\\
&\quad\quad\quad+ \sum_{k = i+1}^N \sum_{l = k+1}^N\comm{h_{ij}}{h_{kl}}\\
&=\sum_{l=j+1}^N \comm{h_{ij}}{h_{il}}\\
&\quad\quad\quad+ \sum_{k = i+1}^N \sum_{l = k+1}^N\comm{h_{ij}}{h_{kl}}\;,\end{split}
\end{equation}
where we have separated the sum for $ L> K $ in two contributions: those where the first index of $K$ is the same as the first index of $L$ and those for which the first index of $L$ is greater than the first index of $K$.
The last contribution can be simplified by splitting the sum around the index $j$ and realizing that, for the commutator $\comm{h_{ij}}{h_{kl}}$ to be different from zero, at least one index in $(k,l)$ needs to match an index in $(i,j)$. The result reads
\begin{equation}
\begin{split}
\sum_{k=i+1}^N \sum_{l=k+1}^N \comm{h_{ij}}{h_{kl} } &= \sum_{k=i+1}^{j-1} \comm{h_{ij}}{h_{kj}} \\
&\quad\quad\quad+ \sum_{l=j+1}^N \comm{h_{ij}}{h_{jl}}\;.
\end{split}
\end{equation}
The Trotter error over a small time step $dt$ thus reads
\begin{equation}
\begin{split}
\varepsilon_1(dt) \leq \frac{dt^2}{2} \sum_{i<j}^N  &\left\|\sum_{k=i+1}^{j-1} \comm{h_{ij}}{h_{kj}}\right.\\
&\quad+\left.\sum_{l=j+1}^N\left( \comm{h_{ij}}{h_{il}} +
 \comm{h_{ij}}{h_{jl}}  \right)  \right\| \,.
\end{split}
\label{Comm_error1}
\end{equation}
The commutators between different two-body Hamiltonians can be computed straightforwardly as
\begin{equation}
\begin{split}
\comm{h_{ij}}{h_{ik}} &= J_{ij} J_{ik} \comm{\boldsymbol{\sigma}_i \cdot \boldsymbol{\sigma}_j}{\boldsymbol{\sigma}_i \cdot \boldsymbol{\sigma}_k} \\
&= 2i J_{ij}J_{ik} \boldsymbol{\sigma}_i \cdot (\boldsymbol{\sigma}_j \wedge \boldsymbol{\sigma}_k)\;,
\end{split}
\end{equation}
where we used $\boldsymbol{a}\wedge\boldsymbol{b}$ to denote the standard cross product in three dimensions. Using the cyclic permutation equivalence of the cross product, the fact that the coupling matrix $J_{ij}$ is positive and the bound $\norm{\boldsymbol{\sigma}_i \cdot (\boldsymbol{\sigma}_j \wedge \boldsymbol{\sigma}_k)} \leq 4$ for all $i,j,k \in [1,N]$ one arrives at
\begin{equation}
    \varepsilon_1(dt) \leq 4dt^2 \sum_{i<j}^N J_{ij}\norm{   \sum_{l=j+1}^N (J_{il}- J_{jl}) + \sum_{k=i+1}^{j-1} J_{kj} }\,.
    \label{errore_Trotter1}
\end{equation}
For a specific choice of angular distributions the sums can be computed straightforwardly. However, in order to obtain a general bound on the error we can introduce $\Theta \coloneqq \max_{i,j}\left[1-\cos(\theta_{ij})\right]$ and obtain the upperbound
\begin{equation}
\varepsilon_1(dt) \leq 12 dt^2 \mu^2 \frac{\Theta^2}{N^2} \binom{N}{3}=\mathcal{O}\left(dt^2\mu^2N\right)\;.
\end{equation}
Using a fixed Trotter time step $dt$ and evolving the system until a total time $T$ using $r = T/dt$ steps the total additive error can be bound by:
\begin{equation}
    \varepsilon_1(T) \leq r \varepsilon_1(dt) \leq 12 \frac{T^2}{r}\mu^2 \frac{\Theta^2}{N^2}\binom{N}{3} = \mathcal{O}\bigg(\frac{T^2 \mu^2N}{r}\bigg)\,.
\end{equation}
In order to have a total error less than some target error $\varepsilon_1(T) \leq \epsilon$ we need a number of steps that scales still linearly with $N$:
\begin{equation}
    r_1 \leq 12 \frac{T^2 \mu^2 \Theta^2}{\epsilon N^2} \binom{N}{3}  = \mathcal{O}\bigg(\frac{T^2 \mu^2 N }{ \epsilon}\bigg)\,.
\end{equation}

\section{Second order Trotter error for the interaction Hamiltonian}
\label{AppendixC}
We present here the extension of the analysis presented in Appendix~\ref{AppendixA} above to the second order Trotter formula. The approximation to the propagator $U_2(dt)$ now reads
\begin{equation}
    U_2(dt) \approx \mathcal{L}_2(dt) = \prod_{L=\Gamma}^1e^{-i\frac{dt}{2}h_{L}} \prod_{K=1}^{\Gamma} e^{-i\frac{dt}{2}h_{K}} \;,
\end{equation}
where we used the multi-index notation $K=(i,j)$, $L=(k,l)$ and $\Gamma=N(N-1)/2$ as before.
Using the result of Proposition 10 from Ref.~\cite{childs2021theory} one can bound the second order Trotter error by:
\begin{equation}
\begin{split}
    \varepsilon_2(dt) \leq& \frac{dt^3}{12} \sum_{K}^{\Gamma} \norm{\sum_{L>K}^{\Gamma} \sum_{M>K}^{\Gamma} \comm{h_L}{\comm{h_M}{h_K}}} \\
    &+\frac{dt^3}{24} \sum_K^{\Gamma} \norm{\sum_{L>K}^{\Gamma} \comm{h_K}{\comm{h_K}{h_L}}}\;.
\end{split}
    \label{trotter_error2}
\end{equation}
In order to bound the second term, we use a similar procedure to the one adopted in the first order case by expanding the sums and keeping contributions $\comm{h_{ij}}{\comm{h_{ij}}{h_{kl}}}$ with one of the $(k,l)$ indices matching one of the $(i,j)$ indices. The bound can be found by using then the expression for the nested commutator
\begin{equation}
\begin{split}
    \comm{h_{ij}}{\comm{h_{ij}}{h_{ik}}}& = J_{ij}^2 J_{ik} \comm{\boldsymbol{\sigma}_i \cdot \boldsymbol{\sigma}_j}{\comm{\boldsymbol{\sigma}_i \cdot \boldsymbol{\sigma}_j}{\boldsymbol{\sigma}_i \cdot \boldsymbol{\sigma}_k}}\\
    &= -4J_{ij}^2 J_{ik} \boldsymbol{\sigma}_i \cdot (\boldsymbol{\sigma}_j \wedge (\boldsymbol{\sigma}_j \wedge \boldsymbol{\sigma}_k))\,,
\end{split}
\end{equation}

together with the bound $\norm{\boldsymbol{\sigma}_i \cdot (\boldsymbol{\sigma}_j \wedge (\boldsymbol{\sigma}_j \wedge \boldsymbol{\sigma}_k))} \leq 8$. The results for the second term in Eq. \eqref{trotter_error2} reads
\begin{equation}
\frac{dt^3}{24} \sum_K^{\Gamma} \norm{\sum_{L>K}^{\Gamma} \comm{h_K}{\comm{h_K}{h_L}}}\leq 4dt^3\frac{\mu^3\Theta^3 }{N^3}\binom{N}{3}\;.
\end{equation}
To determine an upper bound for the first term in Eq. \eqref{trotter_error2} we used the same procedure for both sums inside the norm. In this way four terms are obtained, each of which can be bounded with triangular inequalities by estimating the number of non-zero terms. After some calculations the final result reads
\begin{multline}
    \frac{dt^3}{12} \sum_{K}^{\Gamma} \norm{\sum_{L>K}^{\Gamma} \sum_{M>K}^{\Gamma} \comm{h_L}{\comm{h_M}{h_K}}}\\
    \leq \frac{dt^3}{12}32\frac{\mu^3}{N^3}\Theta^3\left(21\binom{N}{4}+6\binom{N}{3}\right)\,.\\
\end{multline}
So the total error reads
\begin{equation}
\begin{split}
\varepsilon_2(dt) &\leq dt^3 \frac{\mu^3 \Theta^3}{N^3} \bigg[20 \binom{N}{3} + 56 \binom{N}{4} \bigg]\\ 
&= \mathcal{O}\bigg( dt^3 \mu^3 N\bigg)\,,
\end{split}    
\end{equation}
showing a linear scaling with $N$.
The upper bound error after $r = T/dt$ steps is
\begin{equation}
\begin{split}
\varepsilon_2(T) &\leq \frac{T^3}{r^2} \frac{\mu^3 \Theta^3}{N^3} \bigg[20 \binom{N}{3} + 56 \binom{N}{4} \bigg]\\
&= \mathcal{O}\bigg( \frac{T^3}{r^2} \mu^3 N\bigg) \,.
\end{split}
\end{equation}
The total Trotter steps $r$ needed to evolve up to $T$ keeping the total error under $\epsilon$ scales as
\begin{equation}
\begin{split}
r_2 &\leq \frac{(T \mu \Theta)^{3/2}}{\sqrt{\epsilon}N^{3/2}} \sqrt{20 \binom{N}{3} + 56 \binom{N}{4}}\\ 
&=\mathcal{O}\bigg(\frac{T^{3/2} \mu^{3/2}\sqrt{N}}{\sqrt{\epsilon}}\bigg)\,.    
\end{split}    
\end{equation}

\end{document}